\documentclass[prb,reprint,notitlepage,amsmath,amsfonts,amssymb,floatfix]{revtex4-1}
\usepackage{graphicx}
\usepackage{bm}
\usepackage{color}
\usepackage{subcaption}
\usepackage{import}

\begin{document}
\title{A Power Series Approximation in Symmetry Projected Coupled Cluster Theory}

\author{Ruiheng Song}
\email{rs84@rice.edu}
\affiliation{Department of Chemistry, Rice University, Houston, TX 77005-1892}

\author{Thomas M. Henderson}
\affiliation{Department of Chemistry, Rice University, Houston, TX 77005-1892}
\affiliation{Department of Physics and Astronomy, Rice University, Houston, TX 77005-1892}

\author{Gustavo E. Scuseria}
\affiliation{Department of Chemistry, Rice University, Houston, TX 77005-1892}
\affiliation{Department of Physics and Astronomy, Rice University, Houston, TX 77005-1892}
\date{\today}

\begin{abstract}
Projected Hartree-Fock theory provides an accurate description of many kinds of strong correlations but does not properly describe weakly-correlated systems. On the other hand, single-reference methods such as configuration interaction or coupled cluster theory can handle weakly-correlated problems but cannot properly account for strong correlations.  Ideally, we would like to combine these approaches in a symmetry-projected coupled cluster approach, but this is far from straightforward.  In this work, we provide an alternative formulation to identify the so-called disentangled cluster operators which arise when we combine these two methodological strands.  Our formulation shows promising results for model systems and small molecules.
\end{abstract}

\maketitle

\section{Introduction}
Single reference coupled cluster (CC) theory\cite{Paldus1999,Bartlett2007,BartlettShavitt} is considered the `gold standard' for weakly-correlated problems in quantum chemistry.  Although its native $\mathcal{O}(N^6)$ scaling is somewhat formidable, CC with single and double excitations (CCSD)\cite{CCSD} can be applied to very large systems once one exploits locality\cite{Hampel1996,Scuseria1999,Schutz2001,Flocke2004} or uses tensor decompositions\cite{Koch2003,Hohenstein2012,Hohenstein2012b,Benedikt2013,Hummel2017,Schutski2017} to reduce the scaling.

Despite its success in capturing weak correlation, single-reference CC is generally insufficient for strongly-correlated systems in which the single reference is incapable of providing an adequate approximate description of the ground state.  When the number of strongly correlated electrons is not too large, one may overcome this weakness by incorporating higher excitations in the cluster operator (e.g. by adding triple and quadruple excitations to reach CCSDTQ).  This is particularly appealing if one can usefully approximate the higher-order cluster amplitudes rather than solving for them (see, for example, Ref. \onlinecite{Takahashi1984}).  More straightforwardly, one may allow symmetries to break at the mean-field level.  The resulting broken-symmetry mean-field state is usually an unrestricted Hartree-Fock (UHF) determinant which can capture some of the most important features of the relevant strong correlation, and can serve as a useful starting point for standard single-reference CC.  The price paid can be high, however: the broken symmetry can lead to poor prediction of properties, complications with assigning states, and so on.

In order to achieve a better starting point for strong correlation, one can adopt projected Hartree Fock (PHF) theory.\cite{Lowdin1955,Ring1980,Blaizot1985,Schmid2004,PHF}  In PHF, one deliberately breaks %some or all of the 
symmetries of the system, then applies a symmetry projection operator to the broken-symmetry mean-field state.  The end result is a multi-determinantal symmetry eigenstate which is parametrized by a broken-symmetry single determinant.  The underlying mean-field nature of the method means that PHF has mean-field computational scaling.

While PHF can handle many kinds of strong correlation in a black-box manner, the fact that it is a kind of mean-field method means that it is generally unable to describe the weak correlations for which standard correlation techniques are so well suited.  Thus, significant attention has been paid over the past several years to combining PHF with configuration interaction (CI)\cite{Tsuchimochi2016a,Tsuchimochi2016b,Ripoche2017} or even CC.\cite{Duguet2014,Duguet2017,Tsuchimochi2017,Qiu2017,Qiu2018,Tsuchimochi2018,Qiu2019}  The relatively simple nature of the CI wave function means that symmetry-projection of CI is not too complicated.  The CC wave function, however, is far more complicated. Symmetry-adapted alternative coupled cluster methods which incorporate guidance from the PHF wave function have been developed in the past several decades and have shown some promise\cite{Takahashi1984,Piecuch1990,Paldus1991,Piecuch1992,Paldus1996,magoulas2021}, but the exact symmetry-projection of a broken-symmetry CC wave function is not generally feasible.

We are aware of two main approaches one might take to approximately project a CC wave function.  On the one hand, Tsuchimochi and coworkers proposed truncating the exponential of the cluster operator\cite{Tsuchimochi2018} at fourth order, to obtain a balance between computational cost and accurate results. On the other hand, Qiu\cite{Qiu2017} and Duguet\cite{Duguet2014} have independently proposed what we refer to as the disentangled cluster formalism, which we shall explain in more detail below.  The essential feature, however, is that a symmetry-projection operator can be written as a weighted sum of one-body orbital rotation operators, the action of which on a CC state is rather cumbersome. The disentangled cluster formalism writes a rotated CC state as the exponential of a new cluster operator acting on the unrotated reference.  Although formally exact, in practice one must approximate this new cluster operator in some way.  Most of our previous work with the disentangled cluster formalism solved an ordinary differential equation (ODE) to obtain approximate disentangled cluster operators.  In this work, we instead use truncation ideas similar to those of Tsuchimochi and coworkers in combination with the disentangled cluster formulation.  We demonstrate our results for number projected CC in the reduced BCS Hamiltonian, and spin-projected CC in the Hubbard model and a few small molecules.

\section{Projected Coupled Cluster Theory}
Before we discuss our approximation to construct the disentangled cluster operator, we first briefly review symmetry projection and the basic framework of projected CC.  More detailed discussion can be found in, e.g., Ref. \onlinecite{Qiu2019}.

For continuous symmetries such as number or spin, the symmetry projection operator can be written as an integral over symmetry-generated transformations:
\begin{equation}
P = \int d\theta \, w(\theta) \, R(\theta)
\end{equation}
where $R(\theta)$ is a rotation operator (i.e. the exponential of a one-body operator), and $w(\theta)$ is an integration weight which depends on the target symmetry eigenvalue.  Practical calculations discretize the integration.  As a concrete example, we consider number projection, in which the projector can be written as
\begin{equation}
P = \int_0^{2\pi} \mathrm{d}\theta \, \frac{1}{2 \, \pi} \, \mathrm{e}^{-\mathrm{i} \, \theta \, N_0} \, \mathrm{e}^{\mathrm{i} \, \theta \, N}
\end{equation}
where $N$ is the number operator and $N_0$ is the desired particle number.

In projected coupled cluster theory, we write the ground state wave function by acting the projection operator on a broken-symmetry coupled cluster wave function, as
\begin{equation}
|\Psi_\mathrm{PCC}\rangle = P \, \mathrm{e}^U \, |0\rangle = P |\psi_\mathrm{CC}\rangle,
\end{equation}
where $|0\rangle$ is a broken-symmetry determinant, the cluster operator $U$ creates excitations out of $|0\rangle$, and $|\psi_\mathrm{CC}\rangle$ is a broken-symmetry coupled-cluster wave function.  Inserting the projected CC wave function into the Schr\"odinger equation, we obtain
\begin{subequations}
\begin{align}
E &= \frac{\langle 0| H |\Psi_\mathrm{PCC}\rangle}{\langle 0|\Psi_\mathrm{PCC}\rangle} = \frac{\langle 0|H P|\psi_\mathrm{CC}\rangle}{\langle 0|P|\psi_\mathrm{CC}\rangle},
\label{Eqn:PCCEnergy}
\\
0 &= \langle \mu| H - E |\Psi_\mathrm{PCC}\rangle = \langle \mu| \left(H - E\right) \, P |\psi_\mathrm{CC}\rangle,
\label{Eqn:PCCAmps}
\end{align}
\label{Eqn:PCC}
\end{subequations}
where $\langle \mu|$ stands for a set of excited broken-symmetry determinants.

Note that in the absence of the projection operator, Eqns. \ref{Eqn:PCC} reduce to the standard CC equations.  We extract the energy from Eqn. \ref{Eqn:PCCEnergy} and then adjust the amplitudes defining the broken-symmetry cluster operator $U$ by solving Eqns. \ref{Eqn:PCCAmps} for the relevant broken-symmetry determinants $\langle \mu|$ (in projected CCSD, for example, we take $\langle \mu|$ to run over all single- and double-excitations on the reference $\langle 0|$).

We refer the reader to Ref. \onlinecite{Qiu2018} for details on how the amplitude equations are actually solved.  Here, we are chiefly concerned with obtaining the required matrix elements in Eqns. \ref{Eqn:PCC}, for which purpose we use what we call the disentangled cluster approximation.\cite{Duguet2014,Qiu2017,Qiu2018,Qiu2019}

The disentangled cluster approximation is easiest to see by starting with the overlap $\langle 0|P|\psi_\mathrm{CC}\rangle$.  We express this as
\begin{equation}
\langle 0|P|\psi_\mathrm{CC}\rangle = \int \mathrm{d}\theta \, w(\theta) \, \langle 0|R(\theta) \, \mathrm{e}^{U} |0\rangle.
\end{equation}
The key idea is to act the rotation operator to the left on $\langle 0|$, taking advantage of the Thouless theorem\cite{Thouless1960} to write
\begin{equation}
\langle 0|R(\theta) = \langle 0|R(\theta)|0\rangle \, \langle 0| \mathrm{e}^{V_1(\theta)}
\end{equation}
where $V_1(\theta)$ is a single de-excitation operator
\begin{equation}
V_1(\theta) = \sum_{ai} v^i_a(\theta) \, c_i^\dagger \, c_a
\end{equation}
where $i$ and $a$ respectively label orbitals occupied and unoccupied in $\langle 0|$.  We thus have
\begin{equation}
\langle 0|P|\psi_\mathrm{CC}\rangle = \int \mathrm{d}\theta \, w(\theta) \, \langle 0|R(\theta)|0\rangle \, \langle 0| \mathrm{e}^{V_1(\theta)} \, \mathrm{e}^{U} |0\rangle.
\end{equation}
We now act the de-excitation operator on the coupled cluster wave function to the right, and write the result back in coupled cluster form:
\begin{equation}
\mathrm{e}^{V_1(\theta)} \, \mathrm{e}^U |0\rangle = \mathrm{e}^{W(\theta)} |0\rangle
\end{equation}
where $W(\theta)$ consists of a constant ($W_0(\theta)$) and pure excitation operators.  One can use these same basic ideas to evaluate the Hamiltonian overlap and matrix elements between the broken-symmetry coupled cluster wave function and the excited states $\langle \mu|R(\theta)$.

So far, all of this is exact.  The chief difficulty is that even when the cluster operator $U$ is truncated to single and double excitations, the \textit{disentangled} cluster operator $W(\theta)$ contains excitations to all orders and must be in some way approximated.  We might, for example, choose to limit it to include only single and double excitations, discarding its higher-excitation components.  We also must find a convenient way of computing it for each rotation angle $\theta$.  Qiu \textit{et al} proposed integrating a differential equation for $W(\theta)$,\cite{Qiu2017} starting from $W(0) = U$ and truncating the differential equation in some way.  For continuous symmetries such as number or spin, this approach works well; its applicability to discrete symmetries (e.g. a mirror plane spatial symmetry), however, is less certain.

In this work, we propose to follow an alternative scheme for approximately evaluating the disentangled cluster operator.  Although we shall go into more detail shortly, the basic idea is to write a truncated power series expansion of $e^U$ acting on $|0\rangle$, which results in a tractable wave function which can be transformed with $\mathrm{e}^{V_1(\theta)}$ and written in CC form to approximately extract $W(\theta)$.  Note that henceforth we will suppress the explicit gauge-angle--dependence of the disentangled cluster operator $W = W_0 + W_1 + W_2 + \ldots$, where $W_n$ creates $n$-fold excitations when acting on $|0\rangle$.  Note also that this approach was investigated already in Ref. \onlinecite{Qiu2017} but rejected as unsuitable due to slow convergence; we shall show here that in self-consistent calculations, as a practical matter this slow convergence is not so important.

\section{A Series Expansion in Projected CC Theory}
As we have discussed earlier, the approach we investigate here is to assume that the broken-symmetry cluster operator $U$ is in some sense small, so that a power-series expansion in $U$ should make sense.  We begin by writing
\begin{equation}
\mathrm{e}^{V_1} \, \mathrm{e}^{U_1 + U_2 + \ldots} |0\rangle = \mathrm{e}^{W_0 + W_1 + W_2 + \ldots} |0\rangle.
\end{equation}
We find it convenient to separate out $U_1$, writing the left-hand-side of the foregoing equation as
\begin{align}
\mathrm{e}^{V_1} \, &\mathrm{e}^{U_1 + U_2 + \ldots} |0\rangle
 \\ &= \mathrm{e}^{V_1} \, \mathrm{e}^{U_2 + \ldots} \, \mathrm{e}^{-V_1} \, \mathrm{e}^{V_1} \, \mathrm{e}^{U_1} |0\rangle.
\nonumber
\end{align}
From the Thouless theorem, we see that
\begin{equation}
\mathrm{e}^{V_1} \, \mathrm{e}^{U_1} |0\rangle = \mathrm{e}^{W_0^{(1)} + W_1^{(1)}} |0\rangle
\end{equation}
where the notation $W_n^{(k)}$ indicates the portion of $W_n$ which arises from $U_k$.  This means that
\begin{equation}
\mathrm{e}^{W_0 + W_1 + W_2 + \ldots} |0\rangle = \mathrm{e}^{\bar{U}} \, \mathrm{e}^{W_0^{(1)} + W_1^{(1)}} |0\rangle
\end{equation}
where $\bar{U}$ is the similarity-transformation of the non-singles part of $U$ with $V_1$:
\begin{equation}
\bar{U} = \mathrm{e}^{V_1} \, \left(U_2 + U_3 + \ldots\right) \, \mathrm{e}^{-V_1}.
\end{equation}
Transforming $\bar{U}$ again using $W_1^{(1)}$ gives us
\begin{equation}
\tilde{U} = \mathrm{e}^{-W_1^{(1)}} \, \bar{U} \, \mathrm{e}^{W_1^{(1)}}
\end{equation}
and we have, finally,
\begin{equation}
\mathrm{e}^{W_0 + W_1 + W_2 + \ldots} |0\rangle = \mathrm{e}^{W_0^{(1)} + W_1^{(1)}} \, \mathrm{e}^{\tilde{U}}|0\rangle.
\end{equation}

At this point, we expand the exponential to some given order $n$ in the power series expansion, and write the resulting wave function in terms of excitation operators only:
\begin{equation}
\sum \frac{1}{k!} \, \tilde{U}^k |0\rangle = C_0 \, \left(1 + \sum_{k>=1} C_k\right) |0\rangle
\end{equation}
where $C_k$ is a $k$-fold excitation operator; one can then write this configuration-interaction--like wave function in coupled-cluster form so that we obtain
\begin{subequations}
\begin{align}
\mathrm{e}^{W_0} &= C_0 \, \mathrm{e}^{W_0^{(1)}},
\\
W_1 &= C_1 + W_1^{(1)},
\\
W_2 &= C_2 - \frac{1}{2} \, C_1^2.
\end{align}
\end{subequations}
Note that Ref. \onlinecite{Qiu2017} provides an alternative formulation which contains the same substance; we find the approach outlined here conceptually cleaner. However, though this approach may be conceptually straightforward, the actual equations are somewhat cumbersome.  Our derivation and implementation were facilitated by using the symbolic algebra package \textit{drudge} and its accompanying automatic code generation facility \textit{gristmill.}\cite{Drudge}

At this point, it is important to be clear why the expansion we discuss above was abandoned.  When the mean-field reference $|0\rangle$ and the rotated state $\langle 0| R(\theta)$ are nearly orthogonal, the cluster operator $V_1$ is large, so that $\tilde{U}$ may be rather large even if $U_2$ is small.  The power series expansion may converge only slowly, and truncating it at low order might lead to sizable errors.  We shall see later that this concern is frequently not so important; essentially, this is because in the integral over gauge angle $\theta$ the energetically most important region is for $\theta \approx 0$ where $V_1$ is small and this order-by-order expansion works well.  Moreover, our experience suggests that in practice the series expansion converges adequately even for large $V_1$ when the $U$ amplitudes are sufficiently small, as is typically the case, particularly in self-consistent calculations where the $U$ amplitudes are optimized in the presence of the projection operator.  That is, the broken symmetry cluster amplitudes in $U$ are generally smaller when obtained in the presence of the projection operator than when obtained without.\cite{Qiu2018}  Essentially, this is because in the absence of the symmetry projector, the broken-symmetry cluster operator must not only add correlation but must also restore symmetry, but the projector frees the cluster operator from needing to accomplish the second task.

\section{Results}
We test the proposed approach to projected coupled cluster theory both for model systems and for small molecules and consider both projection after variation (PAV) in which we first solve broken-symmetry coupled cluster for the $U$ amplitudes and then project, and variation after projection (VAP) which optimizes the $U$ amplitudes in the presence of the projection operator.  In all cases the broken-symmetry mean-field reference is obtained in a variation after projection sense and is simply the broken-symmetry reference of the projected mean-field theory.  A fully self-consistent procedure which optimizes the mean-field reference in the presence of both the projection operator and the coupled cluster correlator would be very interesting, but is beyond the scope of this work.

The exact result from projected CCSD includes $W$ to all orders, is marked as PCCSD, and is generated using a full configuration interaction code.  

\begin{figure*}[t]
\includegraphics[width=\columnwidth]{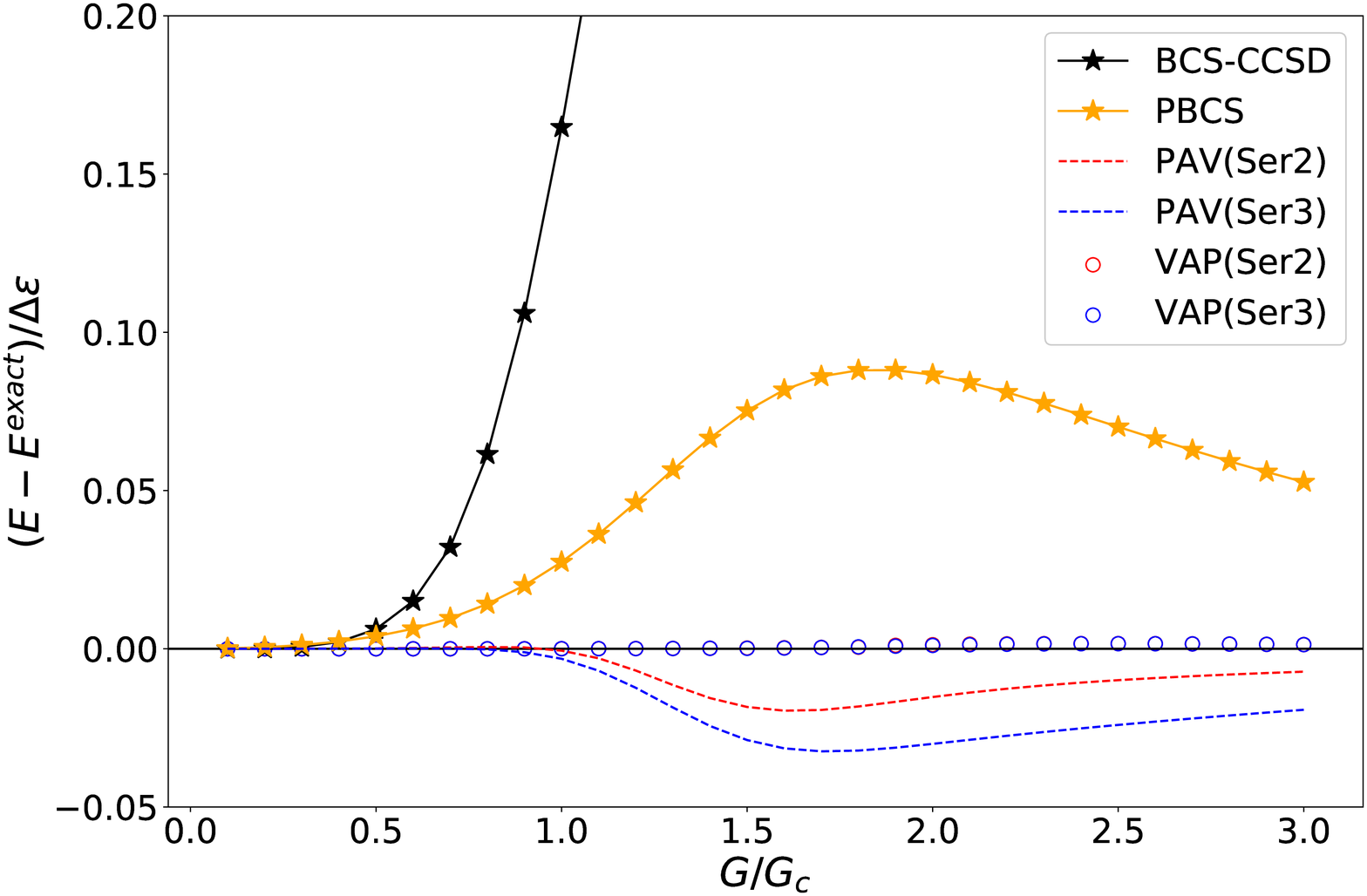} 
\hfill
\includegraphics[width=\columnwidth]{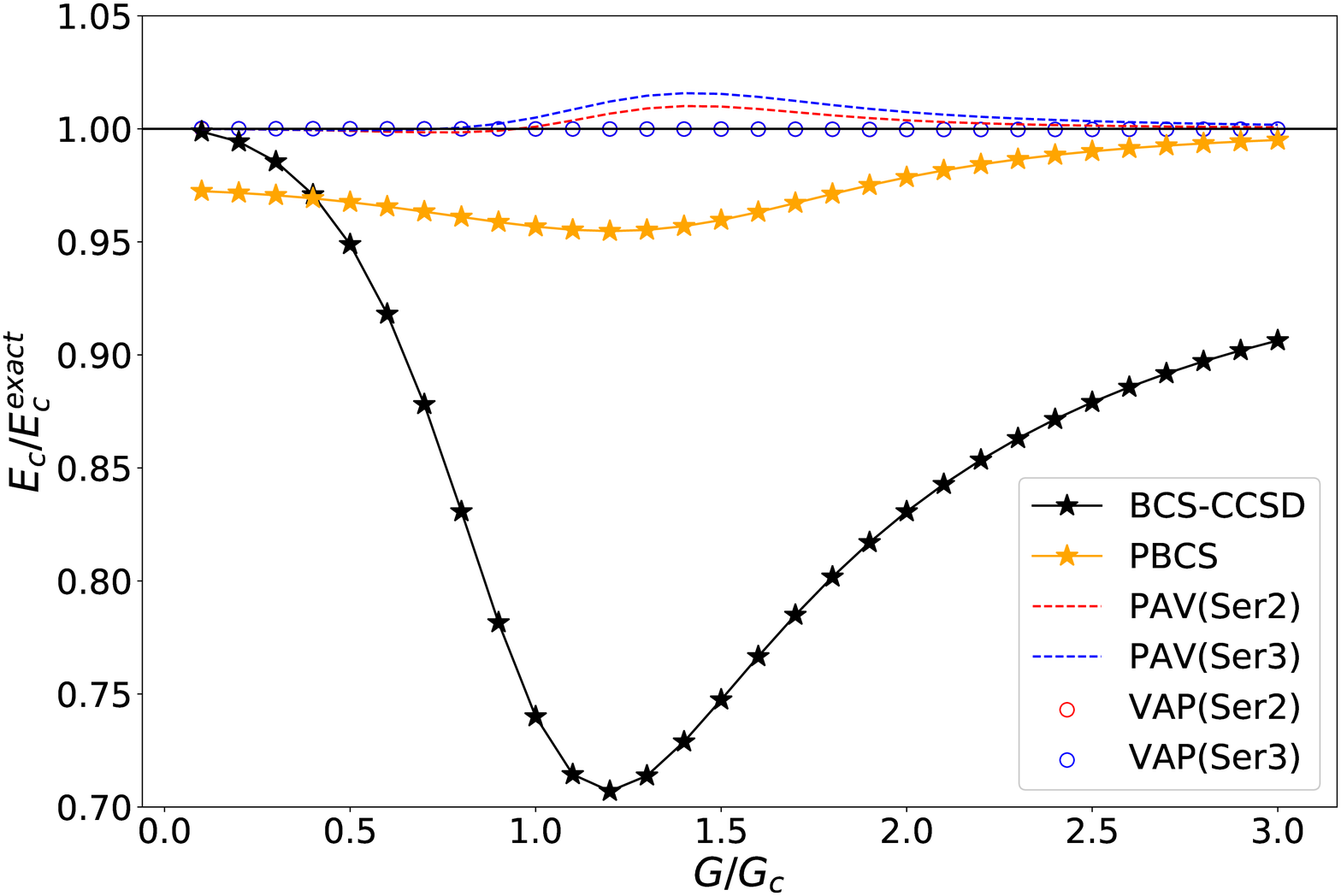}
\caption{Projected coupled cluster in the half-filled 12-site pairing Hamiltonian.  Left panel: error in the total energy.  Right panel: fraction of correlation energy recovered.  Note that the VAP (Ser2) and VAP (Ser3) results are indistinguishable on the scale of the plot.
\label{Fig:Pairing1}}
\end{figure*}

\begin{figure*}
\includegraphics[width=\columnwidth]{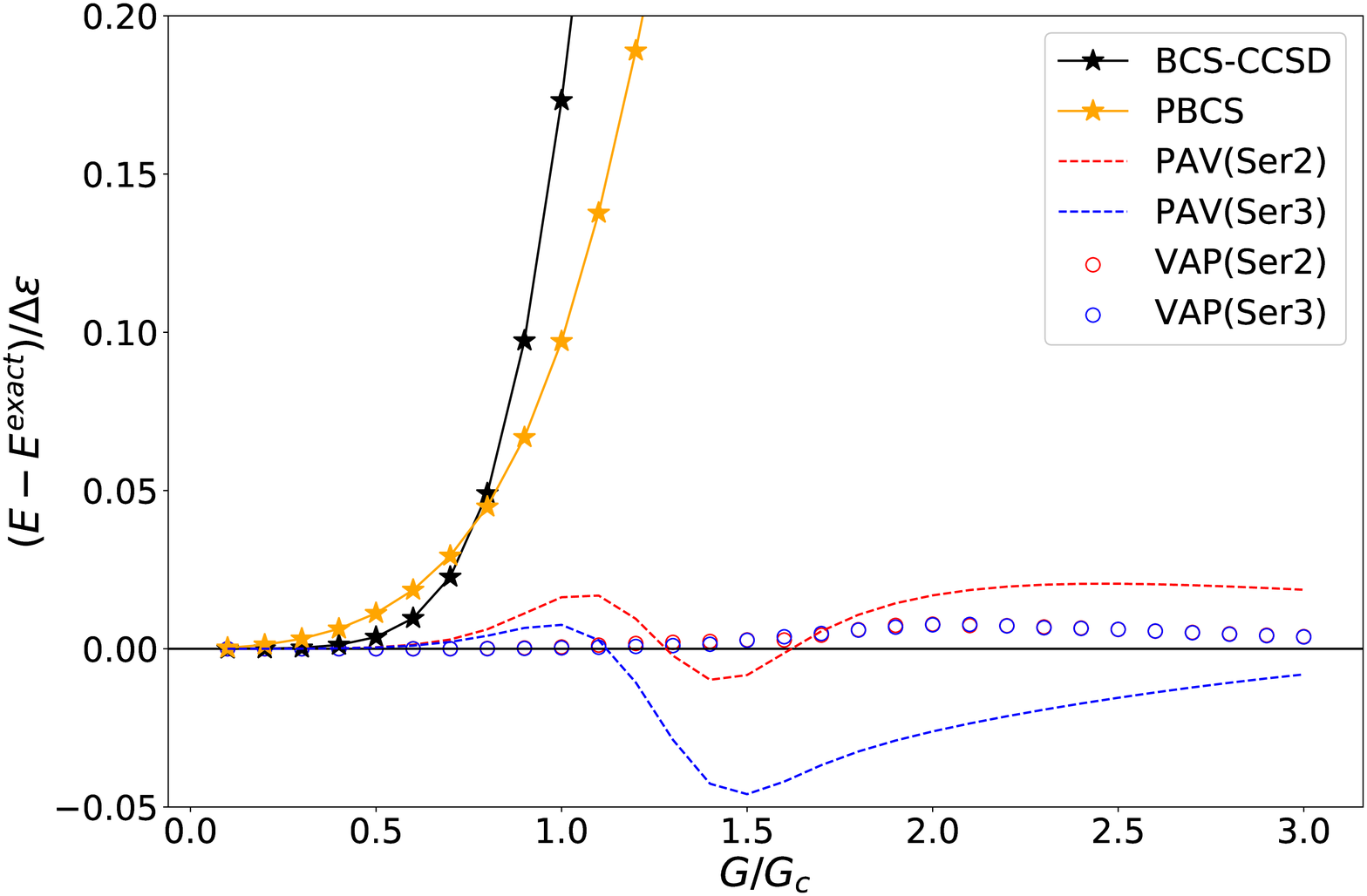} 
\hfill
\includegraphics[width=\columnwidth]{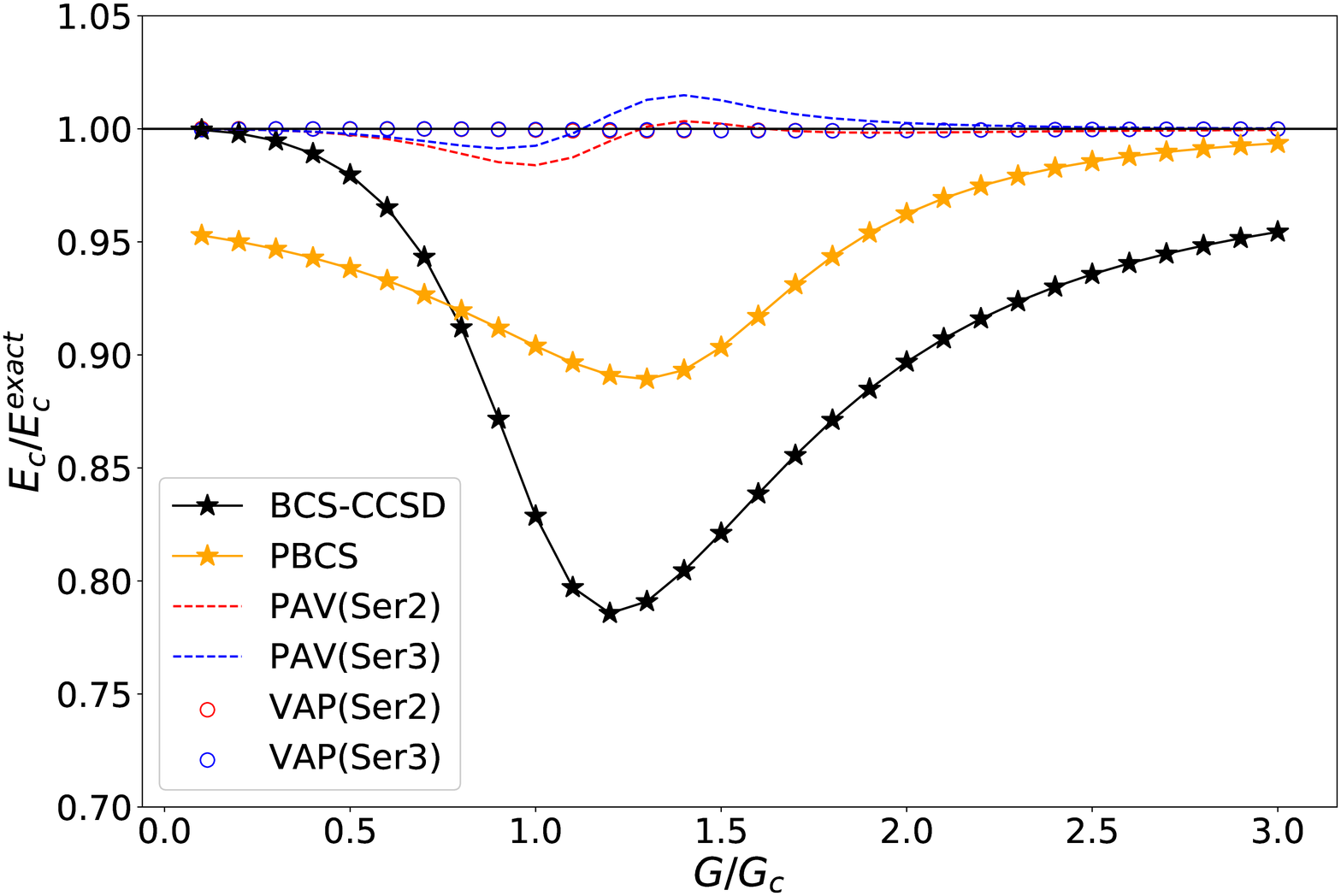}
\caption{Projected coupled cluster in the 40-site pairing Hamiltonian with 10 pairs.  Left panel: error in the total energy.  Right panel: fraction of correlation energy recovered.  Again, VAP (Ser2) and VAP (Ser3) are identical to the eye.
\label{Fig:Pairing2}}
\end{figure*}

Practical calculations require truncating the disentangled cluster operator, and we take $W = W_0 + W_1 + W_2$ so that we neglect triple-excitations and higher in the disentangled cluster operator.  We may then generate $W_0$, $W_1$, and $W_2$ from $U$ and $V_1$ either by the series expansion method or by the ODE approach.  We use `Ser$n$' to indicate that the disentangled cluster operator $W$ truncated at $W_2$ is obtained through $\tilde{U}^n$ and `Ser' for the expansion to all orders; this latter result differs from PCCSD in that PCCSD does not truncate $W$.  We use `ODE2' for the method which Ref. \onlinecite{Qiu2017} refers to SUCCSD(SD).  Briefly, the ODE for $W_2$ depends on $W_3$ and in ODE2, we set $W_3 = 0$ in solving for $W_2$.  With proper optimization, the computational cost for Ser2, Ser3, and ODE2 is $\mathcal{O}(N^6)$ ($\mathcal{O}(N^3)$ for the reduced BCS Hamiltonian which has seniority symmetry).

\subsection{Number-Projected Coupled Cluster}
We begin by considering number projection in the pairing Hamiltonian, given by
\begin{equation}
H = \sum_p \epsilon_p \, N_p - G \, \sum_{pq} \, P_p^\dagger \, P_q.
\end{equation}
Here, the pairing operators can be written as
\begin{align}
P_p^\dagger &= c_p^\dagger \, c_{\bar{p}}^\dagger,
\\
N_p &= c_p^\dagger \, c_p + c_{\bar{p}}^\dagger \, c_{\bar{p}}
\end{align}
where spinorbitals $p$ and $\bar{p}$ are paired.  We take the single-particle levels to be $\epsilon_p = p$.  As the interaction strength $G$ increases, the system becomes strongly correlated, and mean-field spontaneously breaks number symmetry to give a Bardeen-Cooper-Schrieffer\cite{Bardeen1957} (BCS) wave function which we can correlate using number-broken Bogoliubov coupled cluster.\cite{Henderson2014,Signoracci2015}  Although this system is exactly solvable,\cite{Richardson1963,Richardson1964,Richardson1965} conventional coupled-cluster methods are unable to describe it for $G$ much larger than $G_c$, the value at which the mean field spontaneously breaks symmetry.\cite{Dukelsky2003,Henderson2014}

Figure \ref{Fig:Pairing1} shows that projected BCS is accurate for large $G$ but less so for $G \lesssim G_c$, while BCS-CCSD works well for small and very large $G$ (not shown) but fails to describe intermediate $G$ values.  In contrast, projected CCSD is quite accurate across the whole range of interaction strengths, particularly if the cluster amplitudes are optimized in the presence of the projector.  Figure \ref{Fig:Pairing2} shows that these basic features also hold for a larger system away from half-filling.  

\begin{figure}[t]
\includegraphics[width=\columnwidth]{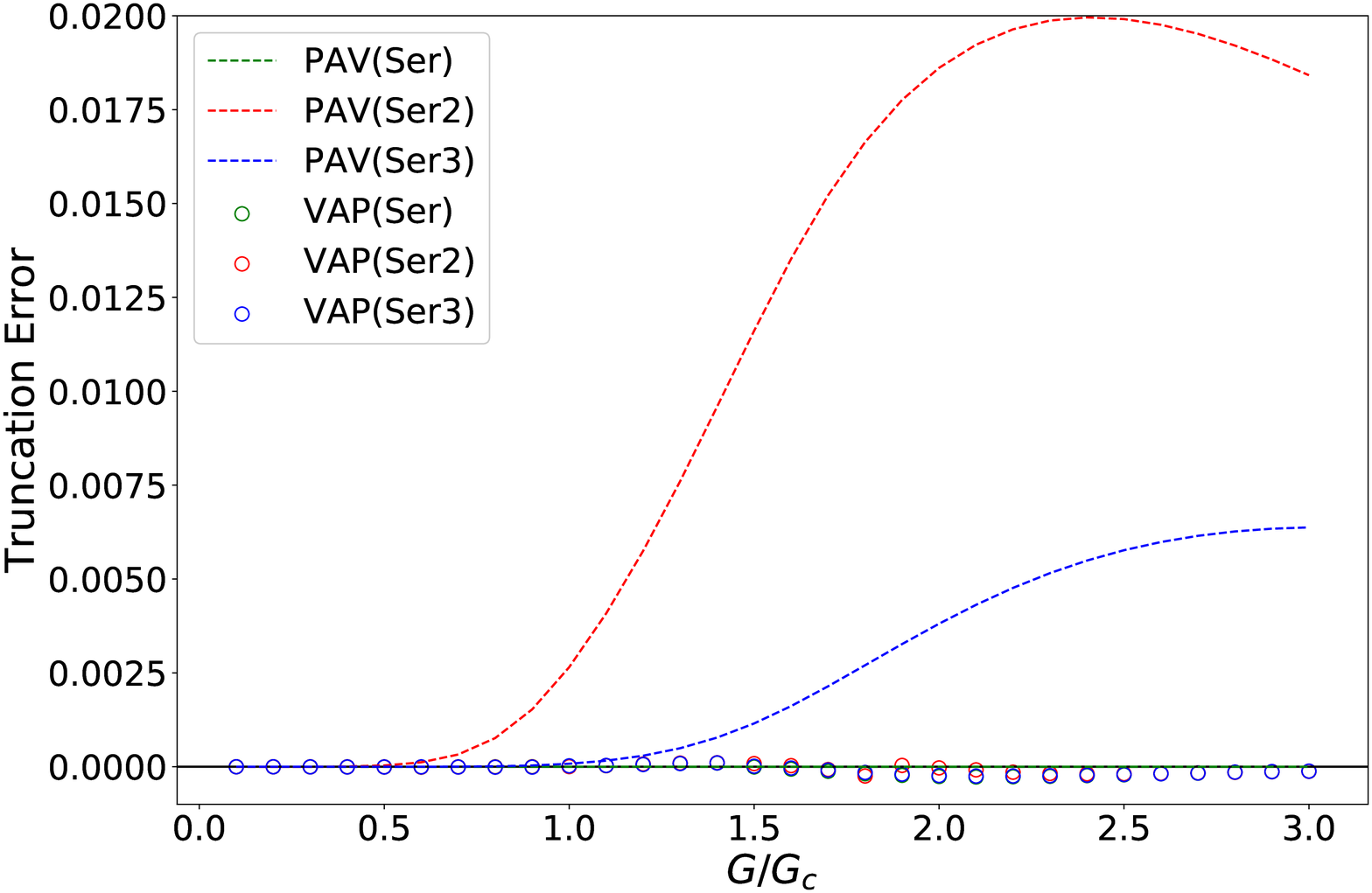} 
\caption{Projected CC errors for the half-filled 12-level pairing model.  Observe that the projected CC energies with series taking through $n=2$ and through $n=3$ are very similar, indicating that the power series expansion gives reasonably well-converged energetics.
\label{Fig:PairingTruncation}}
\end{figure}

Reference \onlinecite{Qiu2019} has already demonstrated that projected CC performs very well for the pairing Hamiltonian.  Our results here are meant to indicate that this is also true if one uses the simple power-series expansion to define the disentangled cluster operators.  

There are three sources of error in our calculations.  The first is the truncation of the broken symmetry cluster operator $U$ to single and double excitations.  We will not concern ourselves with examining the effect of $U_3$ and higher in this work.  The second is the truncation of the disentangled cluster operator at $W = W_0 + W_1 + W_2$, and we define this first kind of truncation error as
\begin{equation}
E_\mathrm{TE1} = E(\mathrm{Ser}) - E(\mathrm{PCCSD}).
\end{equation}
Note that $W_3$ and higher do not directly contribute to the energy, so this first kind of truncation error strictly looks at the effect of $W_3$ and higher on the self-consistent solution for $U_1$ and $U_2$.  The third source of error is the approximate computation of the $W$ amplitudes.  In the series expansion method, this means
\begin{equation}
E_\mathrm{TE2} = E(\mathrm{Ser}n) - E(\mathrm{Ser}).
\end{equation}
We would have a similar result for the ODE.  The total truncation error is the sum of $E_\mathrm{TE1}$ and $E_\mathrm{TE2}$, and can help us evaluate the validity of different approximations.

\begin{figure}
\includegraphics[width=\columnwidth]{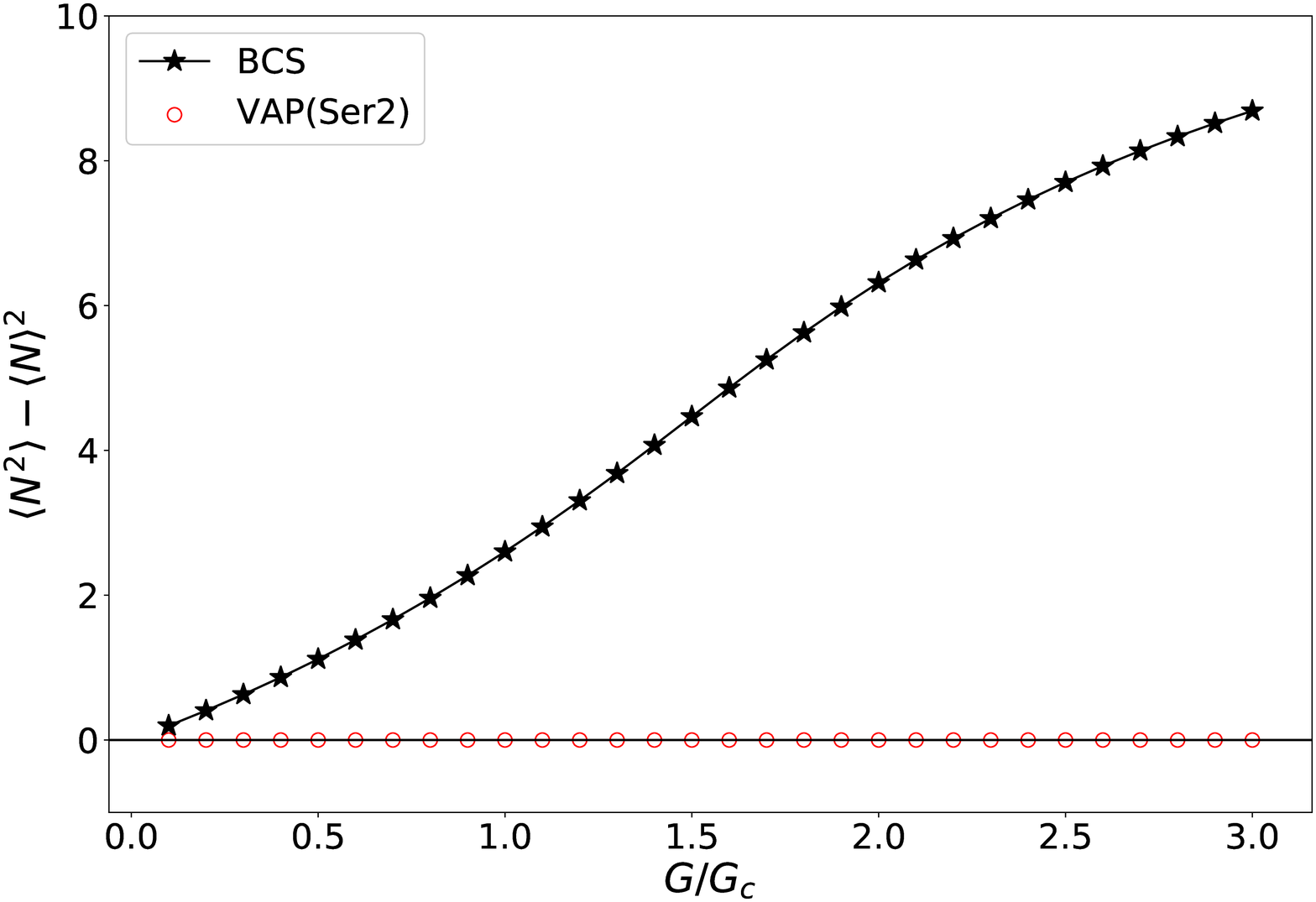} 
\\
\includegraphics[width=\columnwidth]{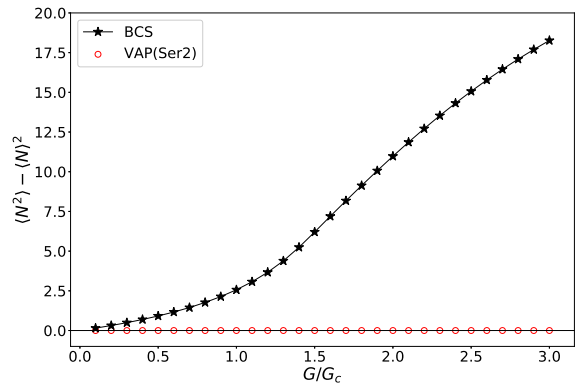}
\caption{Variance in particle number in the pairing Hamiltonian.  Top panel: 6 pairs in 12 levels.  Bottom panel: 10 pairs in 40 levels.
\label{Fig:PairingFluctuations}}
\end{figure}

For the half-filled 12-level system, $E_\mathrm{TE1}$ is almost zero for all $G$, which suggests that the effects of $W_3$ and higher on the converged $U$ amplitudes are essentially negligible and the truncation error mainly arises from approximating $W_0 - W_2$.  We see that even in the PAV(Ser2) case, the truncation error is not large (truncation errors are on the order of $10^{-2}$) and PAV(Ser3) performs much better.  In the VAP cases, the truncation error is negligible after optimizing the $U$ amplitudes.

Note that our results, confirming those of previous work, imply that optimization of the cluster amplitudes plays a vital role in projected coupled cluster theory, so that even a finite-order expansion can lead to remarkable agreement with the exact projected CC result.

Finally, as we have discussed previously,\cite{Qiu2017} truncating the disentangled cluster operator to low-order (e.g. through $W_2$) implies approximating the projection operator, so that the symmetry restoration is imperfect.  Figure \ref{Fig:PairingFluctuations} shows fluctuations $\langle N^2 \rangle - \langle N \rangle^2$ in particle number as a function of $G$.  Even with $W$ truncated and with our low-order power series approximations to the truncated $W$, we see that the symmetry is restored almost perfectly and the fluctuations are essentially zero.

\begin{figure}[t]
\includegraphics[width=\columnwidth]{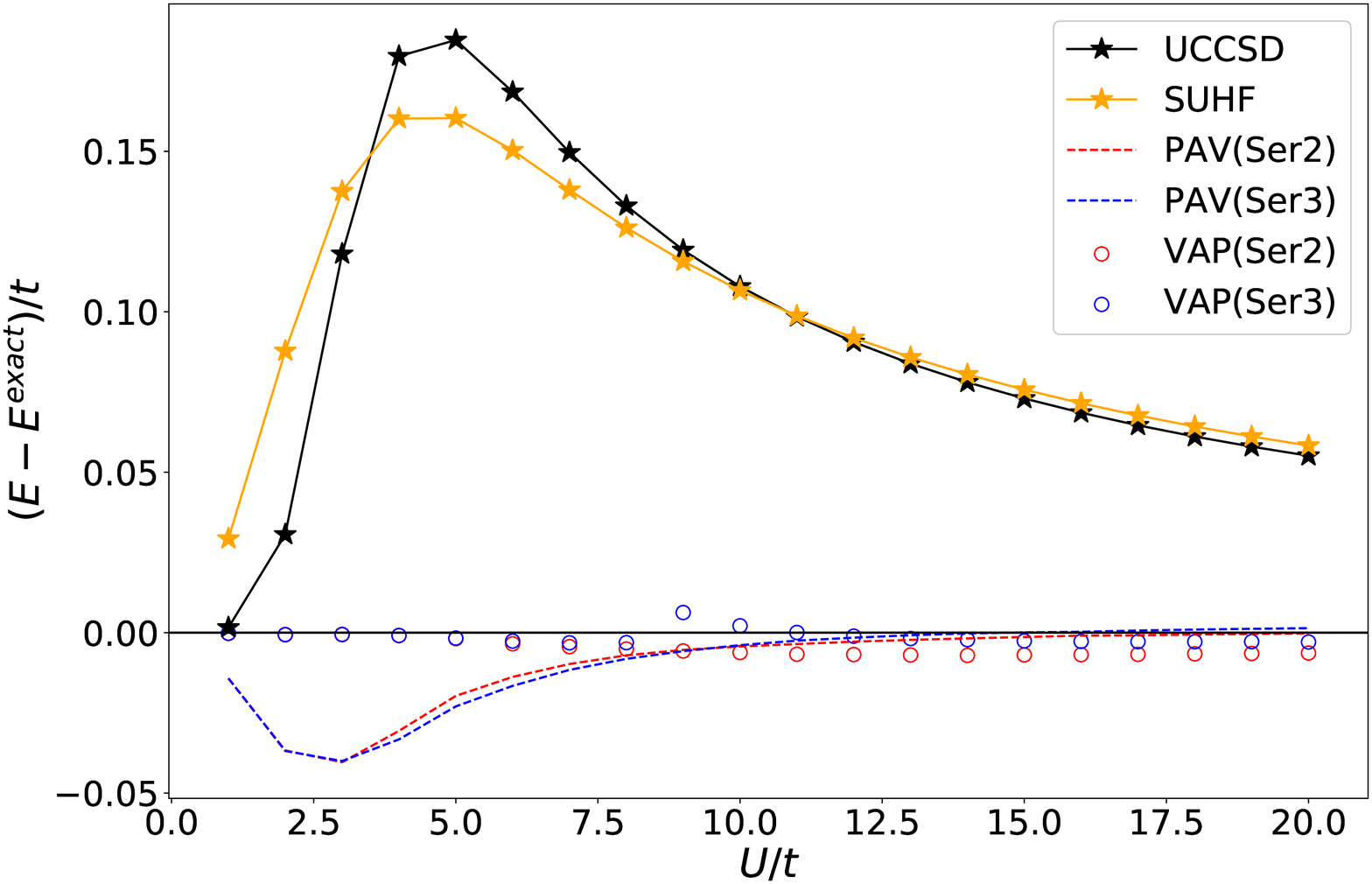} 
\\
\includegraphics[width=\columnwidth]{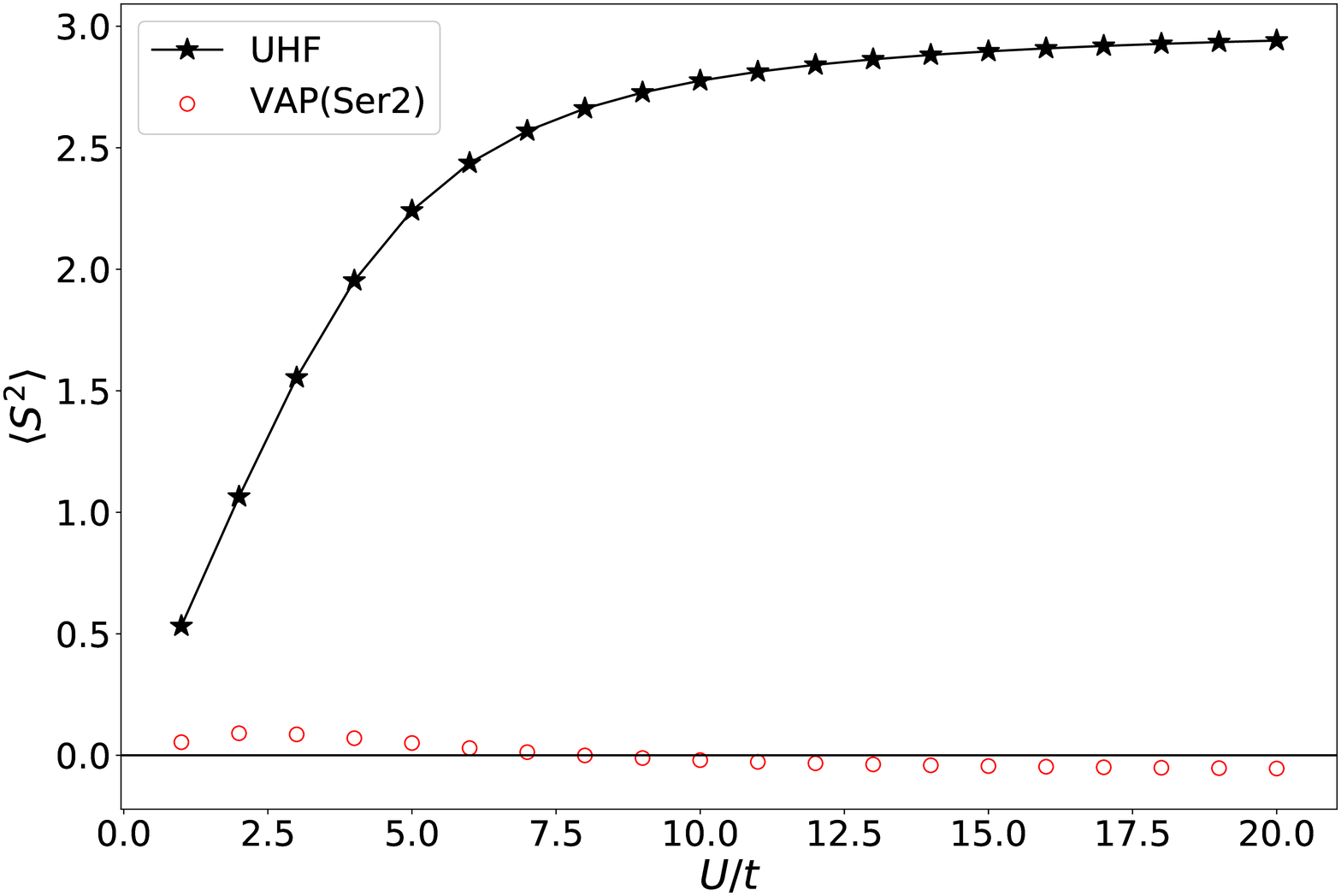}
\caption{Projected CCSD results in the half-filled 6-site Hubbard model.  Top panel: Errors in the total energy.  Bottom panel: Expectation value of $S^2$.
\label{Fig:Hubbard1}}
\end{figure}

\subsection{Spin-Projected Coupled Cluster}
For repulsive two-body interactions, number symmetry breaking is not generally as important (and number is not broken spontaneously at the mean-field level when the interaction is repulsive).\cite{Bach1994}  Instead, spin symmetry breaks spontaneously.  Generally, we are going to be interested only in breaking $S^2$ but not $S_z$ spin symmetry, which at the mean-field level means that we are interested in unrestricted Hartree-Fock (UHF) but not general Hartree-Fock (GHF).  This is simply because most molecules of interest do not have distinct GHF solutions (though some do) while UHF solutions are ubiquitous; they appear in any open-shell molecule and in dissociation of closed-shell molecules to open-shell fragments.

Although we are principally interested in molecular systems, we begin instead with the Hubbard Hamiltonian,\cite{Hubbard1963} a lattice model whose Hamiltonian is
\begin{equation}
H = -t \, \sum_{\langle ij \rangle,\sigma} \left(c_{i,\sigma}^\dagger \, c_{j,\sigma} + \textit{h.c.}\right) + U \, \sum_i n_{i,\uparrow} \, n_{i,\downarrow}
\end{equation}
where the notation $\langle ij \rangle$ indicates we include only sites $i$ and $j$ connected in the lattice.  The first term favors electronic delocalization, while the second penalizes double-occupancy of a lattice site and thus favors electronic localization.  As $U/t$ increases, the Hubbard model becomes more strongly correlated.

Although the Hubbard model contains more interesting physics in two dimensions, we will focus here on the one-dimensional Hubbard model, which can be solved exactly\cite{LiebWu} and which is readily accessible with a full configuration interaction (FCI) code.  Away from half filling, the Hubbard model generally has GHF solutions for large $U/t$, but we are more interested in spin-projected unrestricted CCSD since, as we have noted, molecules tend not to have GHF solutions.  Moreover, spin-projected GHF is already rather cumbersome computationally, and while projected GHF-based CCSD can certainly be implemented, we do not have such a code.  We therefore limit ourselves to the half-filled Hubbard model in this work, as it has UHF but not GHF solutions.  We note, however, that the doped Hubbard model (i.e. away from half filling) is more challenging for projected Hartree-Fock methods, and it is not clear that simple spin-projected CC would be adequate in any event since spin-projected Hartree-Fock has large errors for the doped Hubbard model at large $U/t$.\cite{PhysRevX2015}  Since PHF has significant remaining correlation to capture, we would generally expect the $U$ amplitudes in broken-symmetry CCSD to be large, and our series expansion approach is unlikely to work well in such a case even if projected CCSD is in principle accurate.

\begin{figure}
\includegraphics[width=\columnwidth]{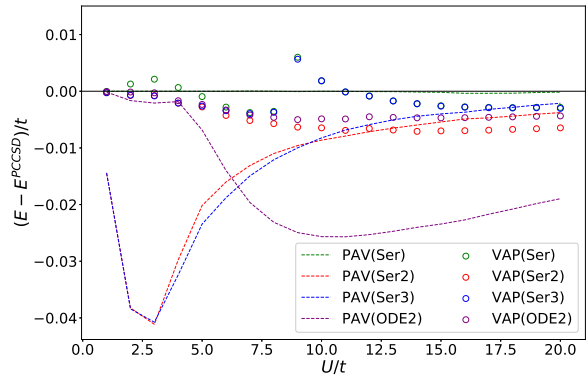} 
\caption{Truncation error in the half-filled 6-site Hubbard model.  Note that PAV(Ser) has essentially zero error for $U/t \lesssim 15$ and sits underneath the $x$ axis.
\label{Fig:HubbardTrunc}}
\end{figure}

All that being said, we show results for the 6-site half-filled Hubbard model with periodic boundary condition in Fig. \ref{Fig:Hubbard1}.  We see that unrestricted CCSD (UCCSD) has relatively large errors and is generally competitive with spin-projected UHF (SUHF).  Errors are greatly reduced by spin projection of the UCCSD.  The result for the VAP PCCSD approach are particularly good; once again, solution of the broken-symmetry cluster amplitudes in the presence of the projector is important for achieving best accuracy.  We have difficulty converging VAP-PCCSD with third-order or higher power series (i.e. Ser(3) and Ser) near $U/t \approx 8$.  This may be due to a linear dependency in the amplitude equations, which is discussed in more detail elsewhere.\cite{Qiu2018,Tsuchimochi2018}  This one problem aside, our power series approach works well here and as Fig. \ref{Fig:HubbardTrunc} shows is quite close to the exact PCCSD at large $U/t$.

Note that, as with the pairing Hamiltonian, the truncation error is much larger for the PAV approach, although interestingly, the PAV approach gives slightly superior total energies for large $U/t$. The large error in PAV(Ser2) and PAV(Ser3) at small to moderate $U/t$ is caused by the large $U$ amplitudes. We notice that, at $U/t=2$, the maximum $U_1$ and $U_2$ are on the order of $10^{-1}$, while at $U/t=20$, the maximum goes down to $10^{-2}$. As one might expect, truncating the series expansion works well when the broken-symmetry cluster amplitudes in $U$ are small, but is less appropriate when the $U$ amplitudes are large.

\begin{figure}[t]
\includegraphics[width=\columnwidth]{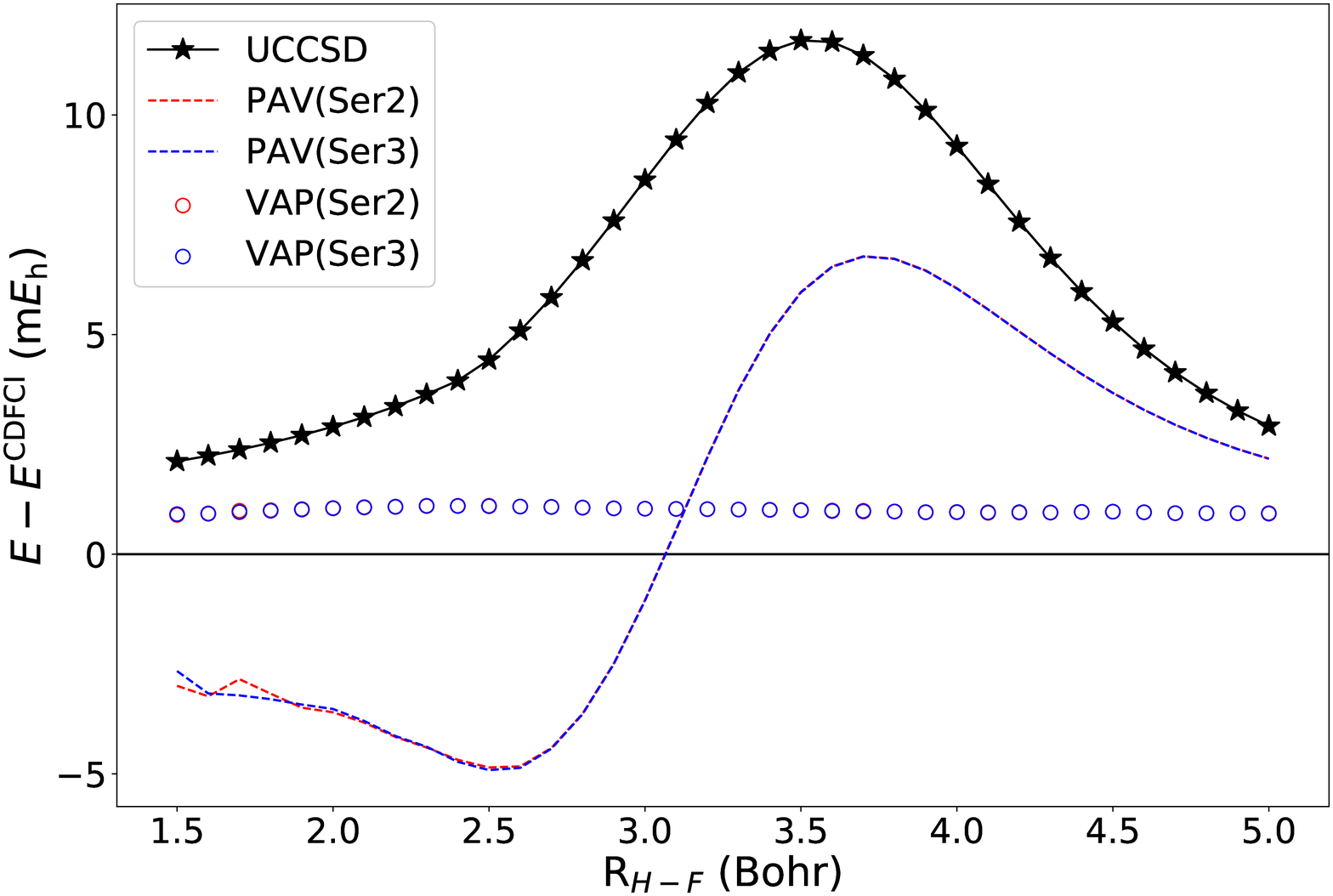} 
\\
\includegraphics[width=\columnwidth]{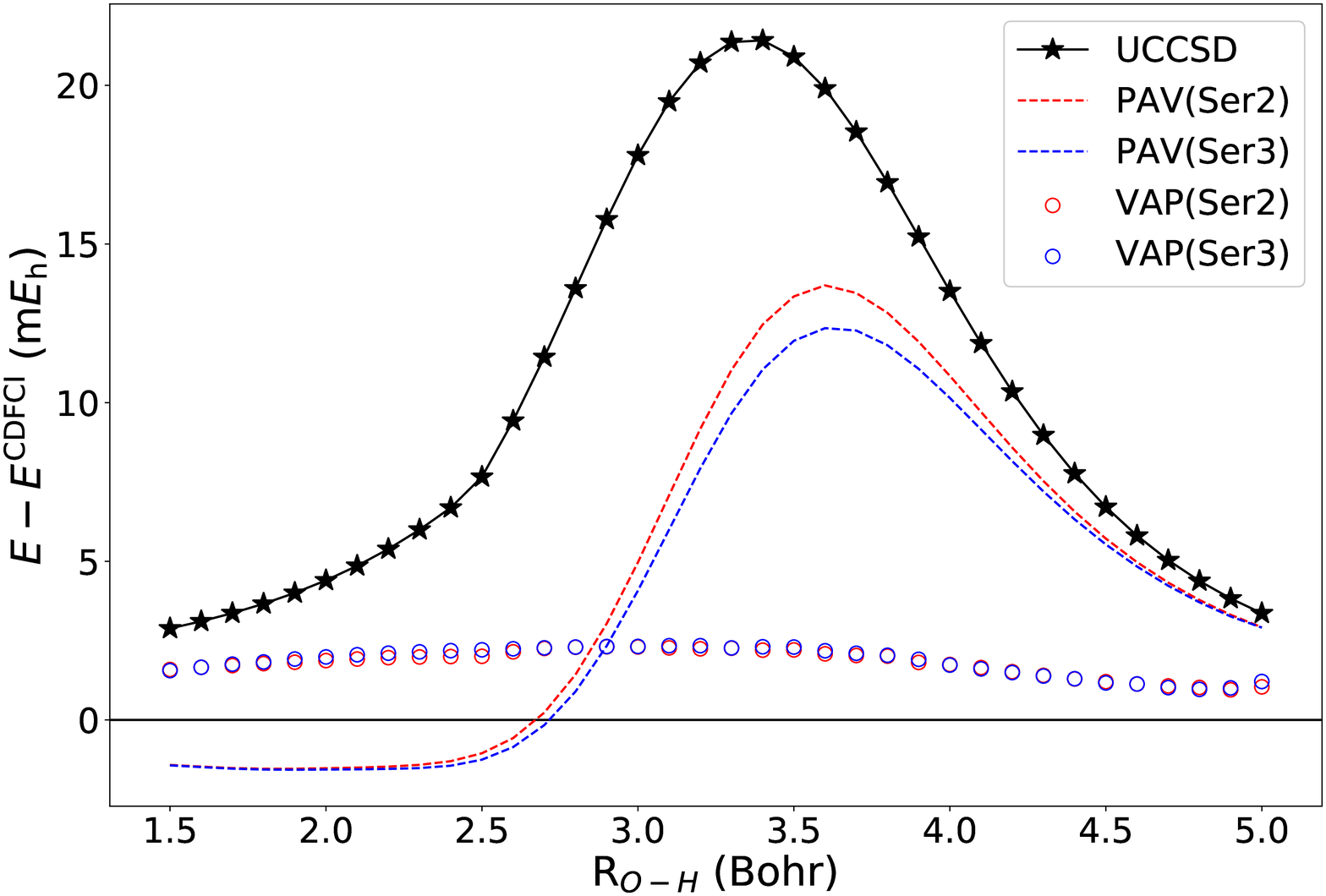}
\caption{Energy errors in molecular dissociations.  Top panel: dissociation of HF.  Bottom panel: Symmetric dissociation of H$_2$O to H + O + H.
\label{Fig:Molecules}}
\end{figure}

We now turn to a few small molecular dissociation curves, all using the cc-pVDZ basis set. We use PySCF\cite{pyscf,pyscf2} and the coordinate-descent full configuration interaction\cite{cdfci} (CDFCI) method as implemented in CDFCI\cite{CDFCI2} to generate benchmark results deemed to be of near FCI quality. We refer these results as $E^{\mathrm{CDFCI}}$. Figure \ref{Fig:Molecules} shows results for the dissociation of HF and the symmetric double dissociation of H$_2$O ($\angle$ HOH = 104.5$^{\circ}$).  We exclude SUHF from these plots since, while it gives qualitatively reasonable dissociation curves it has large total energy error.  Unrestricted CCSD is fairly good, but does not deliver chemical accuracy, as one might expect.  Adding the projection operator in the PAV sense does not significantly improve upon UCCSD and generally overcorrects; in contrast, the VAP-PCCSD is excellent and differs from the reference results by around 1 mE$_\textrm{h}$ for all bond lengths.  The truncation error is presumably small since VAP(Ser2) and VAP(Ser3) give almost identical results everywhere.

\begin{figure}[t]
\includegraphics[width=\columnwidth]{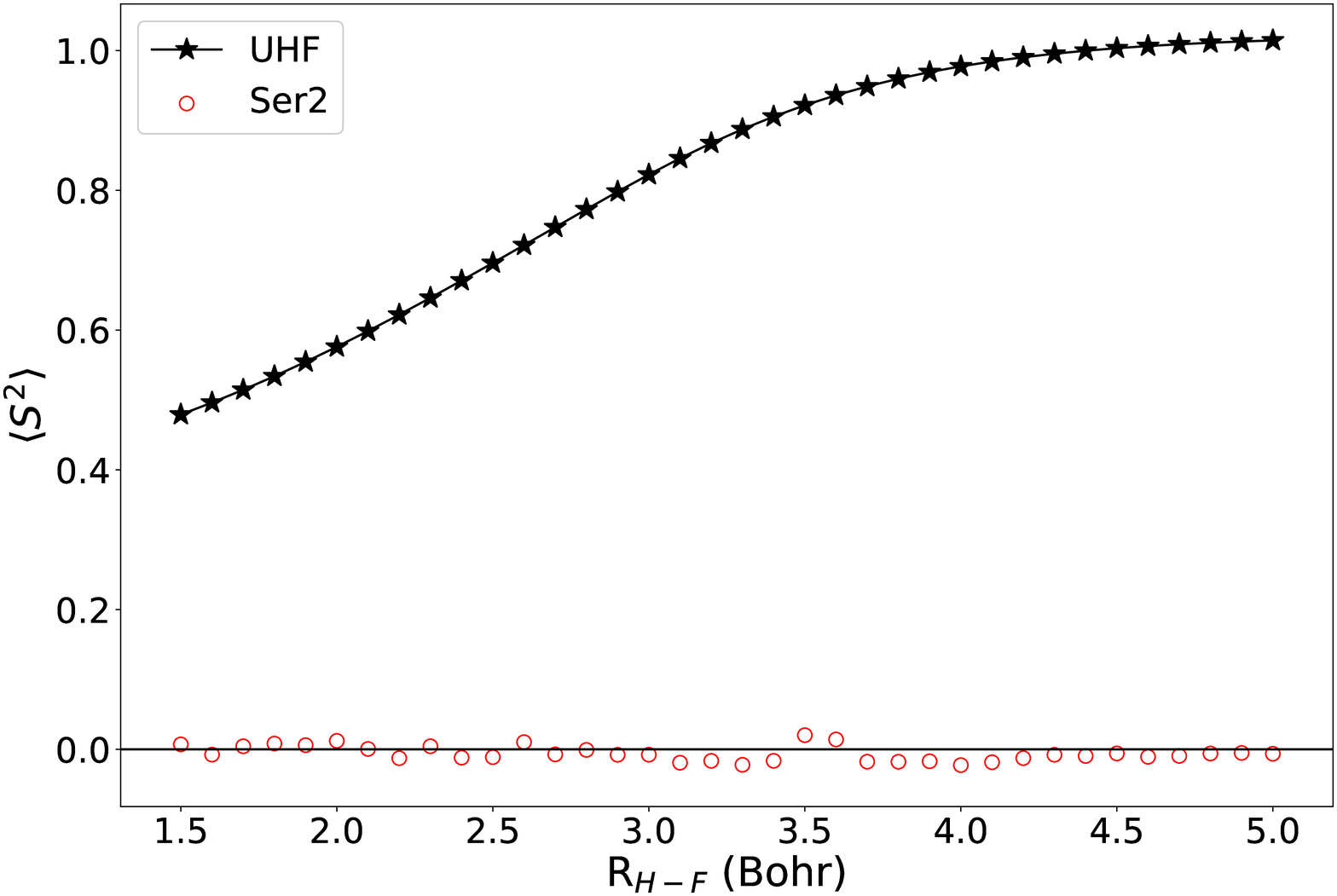} 
\\
\includegraphics[width=\columnwidth]{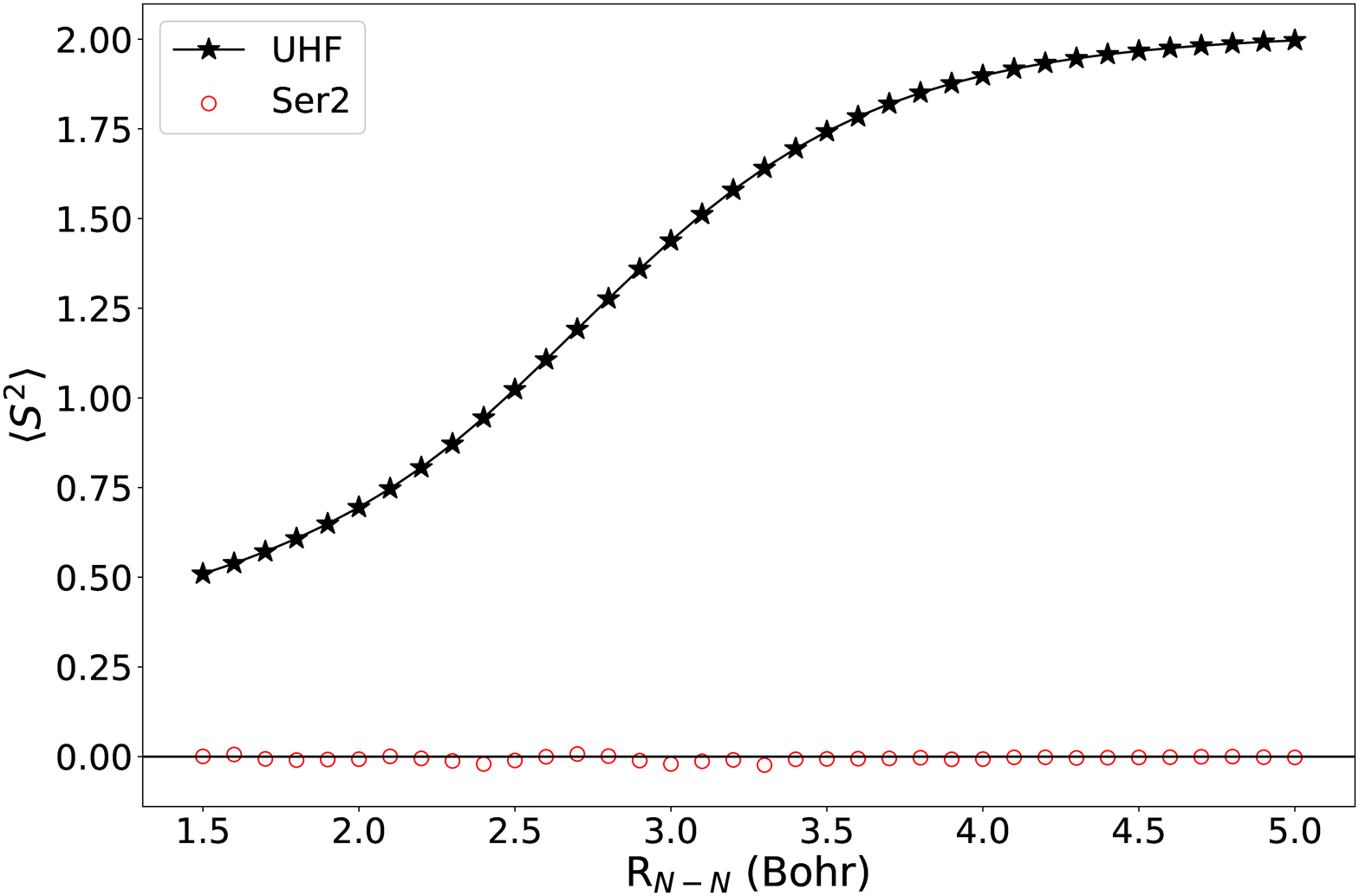}
\caption{The expectation value $\langle S^2 \rangle$ in small molecules.  Top panel: HF.  Bottom panel: H$_2$O.
\label{Fig:S2}}
\end{figure}

These results suggest that VAP(Ser2) is frequently going to be good enough for practical purposes.  Figure \ref{Fig:S2} shows that indeed, such an approach largely eliminates the spin contamination for the small molecules we have considered.  The errors are somewhat larger than in the Hubbard model, and are not entirely systematic, but we think they are small enough given the accurate energetic predictions.

\begin{figure}[t]
\includegraphics[width=\columnwidth]{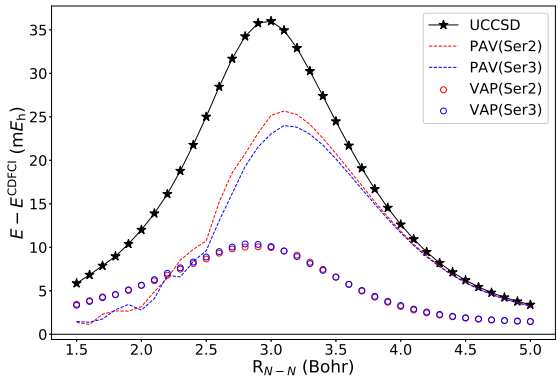} 
\caption{Energy error in N$_2$ dissociation.
\label{Fig:N2}}
\end{figure}

\begin{figure}
\includegraphics[width=\columnwidth]{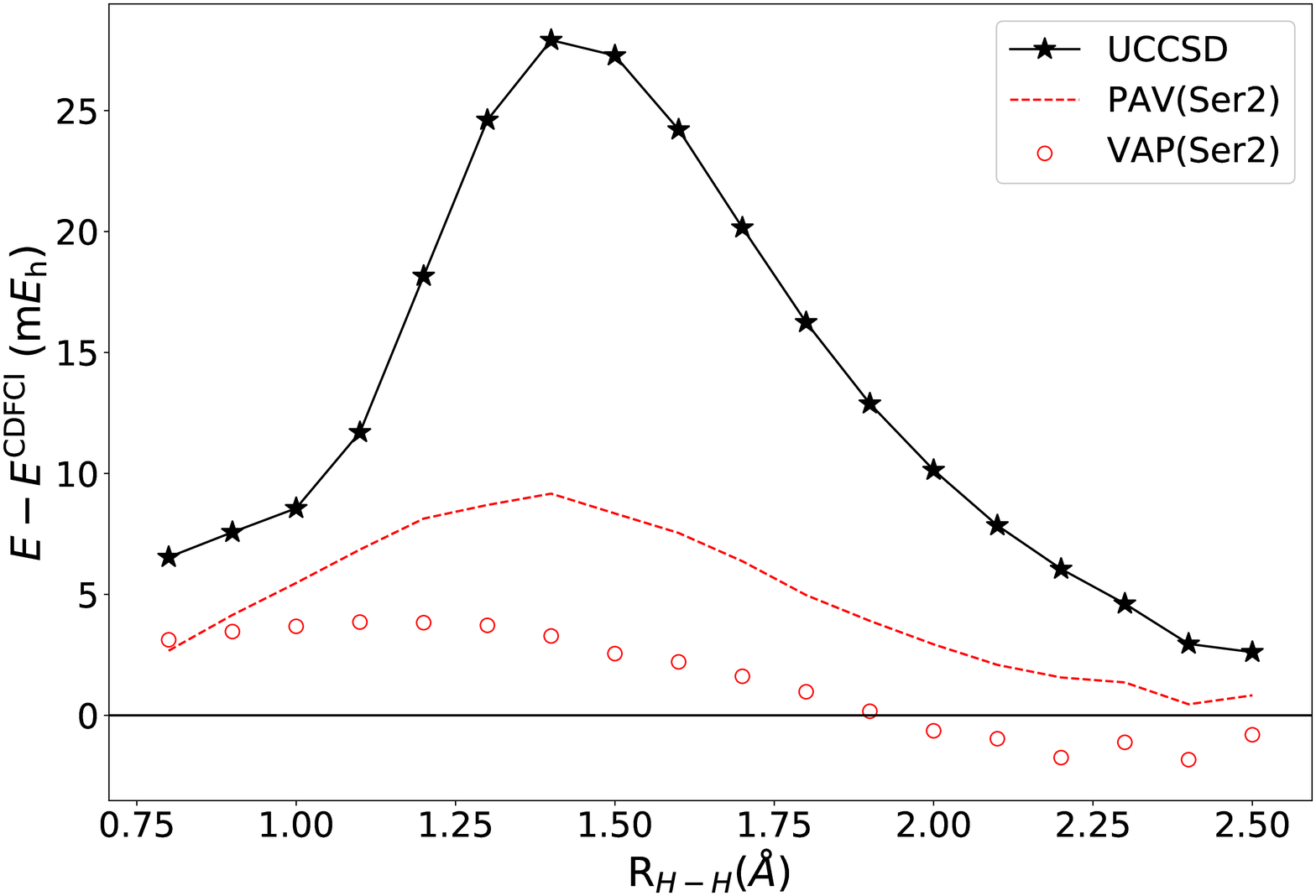} 
\caption{Energy error in H$_{10}$ symmetric dissociation.
\label{Fig:H10}}
\end{figure}

We look at more difficult examples in Figs. \ref{Fig:N2} and \ref{Fig:H10}, which respectively consider dissociation of N$_2$ and of the H$_{10}$ ring; as the latter system is somewhat larger we apply only the Ser2 method.  Although we do less well for these systems, PCCSD still greatly improves upon UCCSD, particularly when done in the VAP manner.

\begin{figure}[t]
\includegraphics[width=\columnwidth]{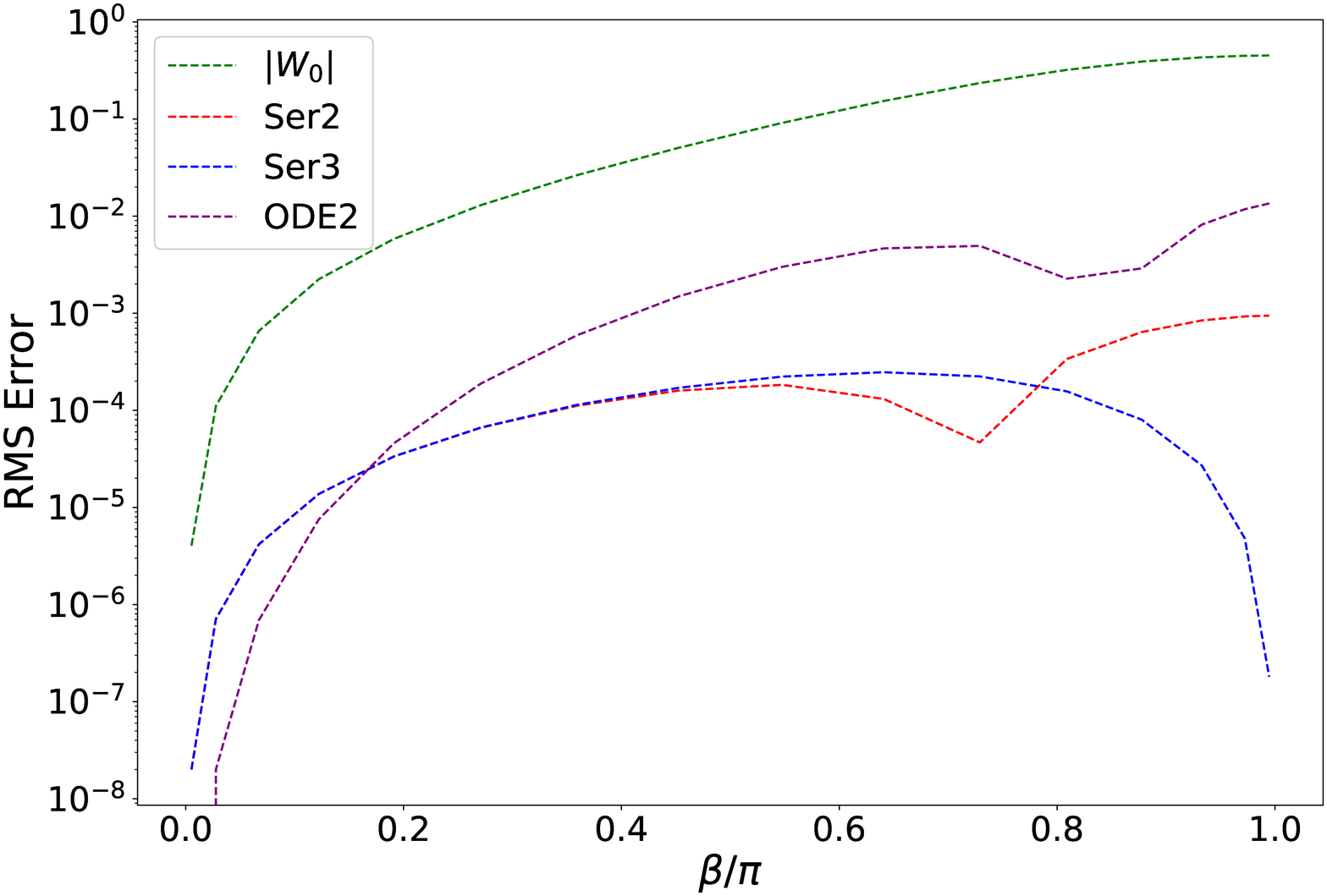} 
\\
\includegraphics[width=\columnwidth]{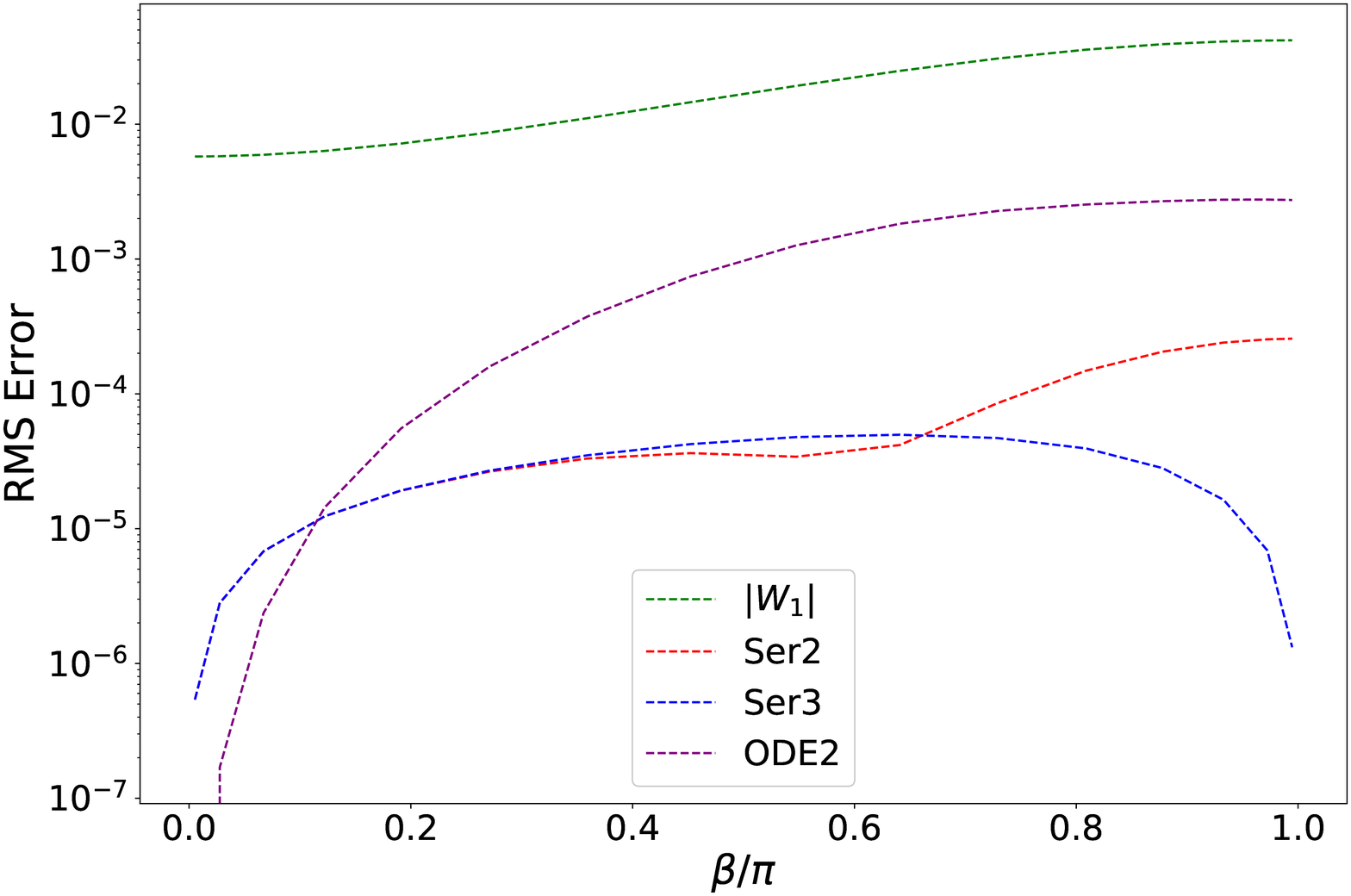}
\\
\includegraphics[width=\columnwidth]{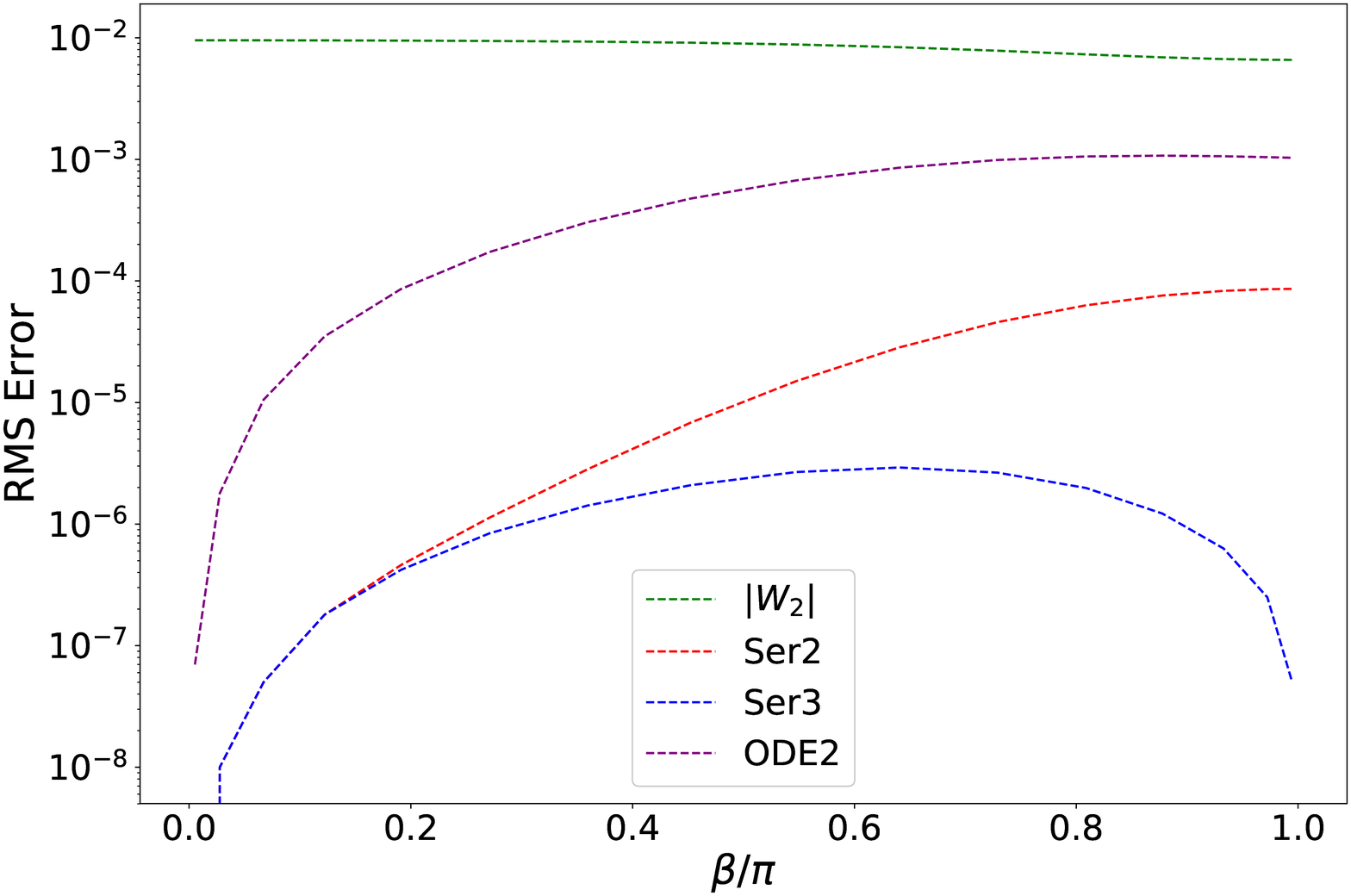}
\caption{RMS error in $W$ from different approximation at $U/t=4$ with exact $U$.  Top panel: Error of $W_0$.  Middle panel: Error of $W_1$. Bottom panel: Error of $W_2$. 
\label{Fig:4PAV}}
\end{figure}

\begin{figure}[t]
\includegraphics[width=\columnwidth]{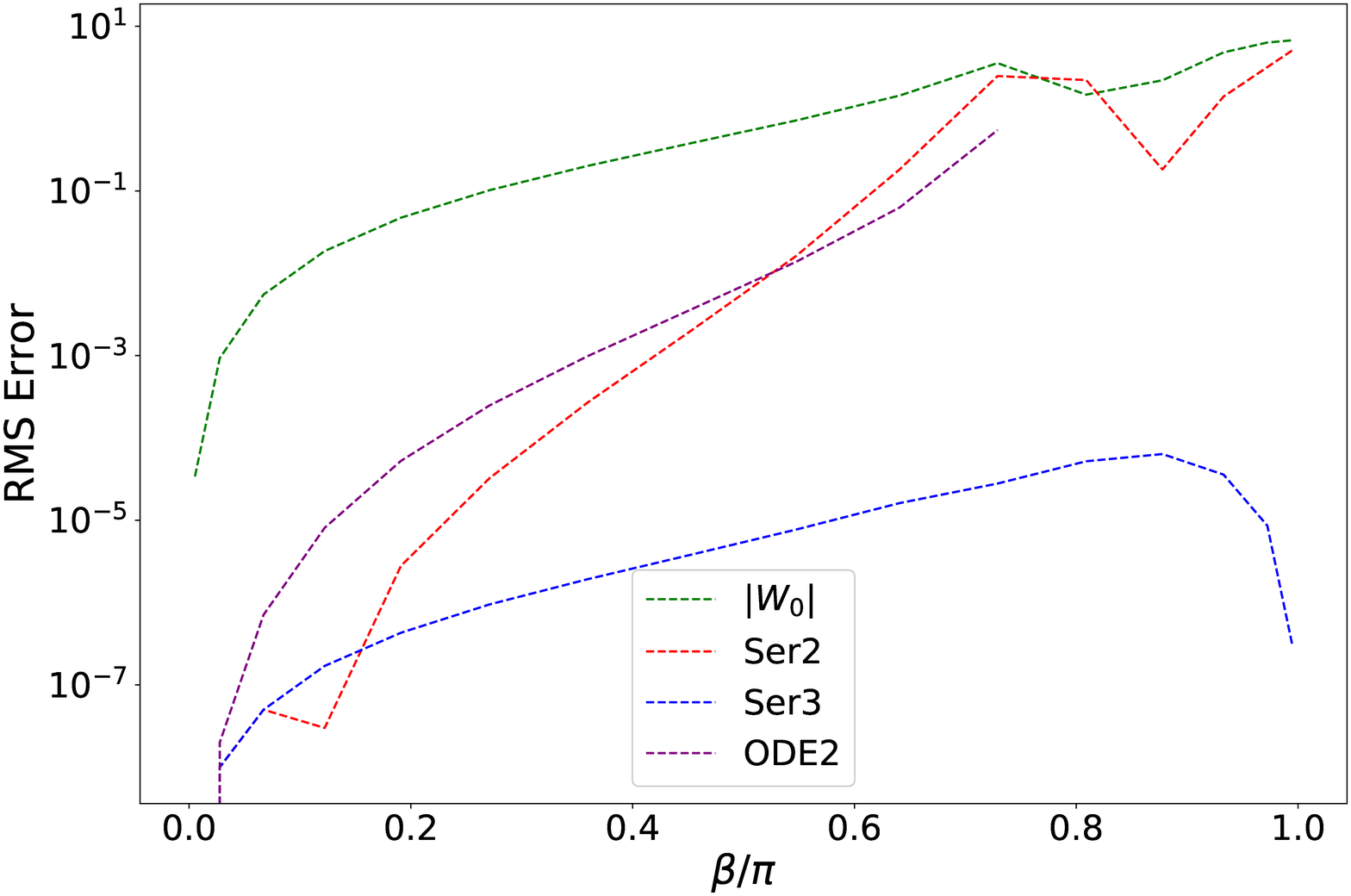} 
\\
\includegraphics[width=\columnwidth]{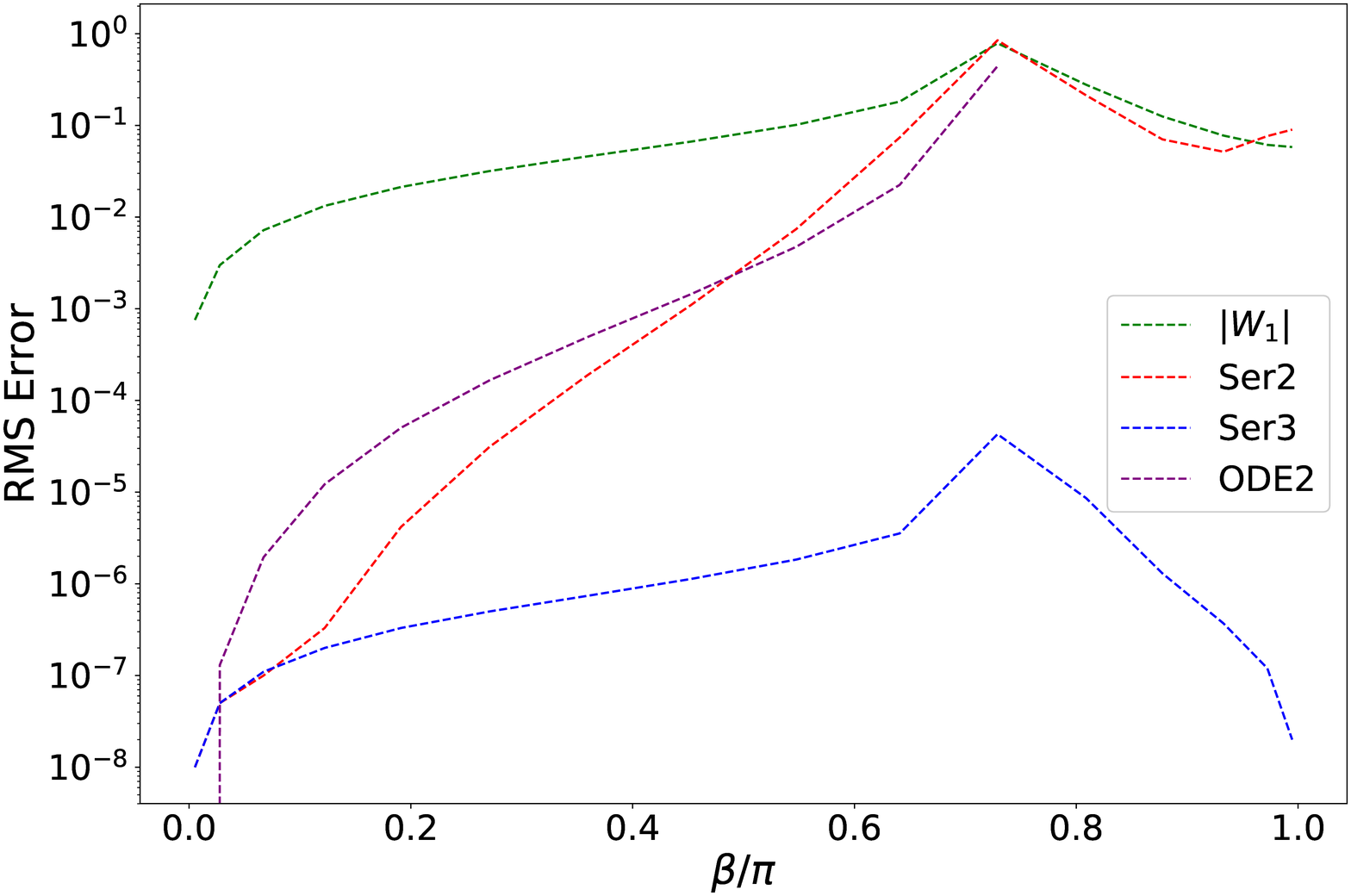}
\\
\includegraphics[width=\columnwidth]{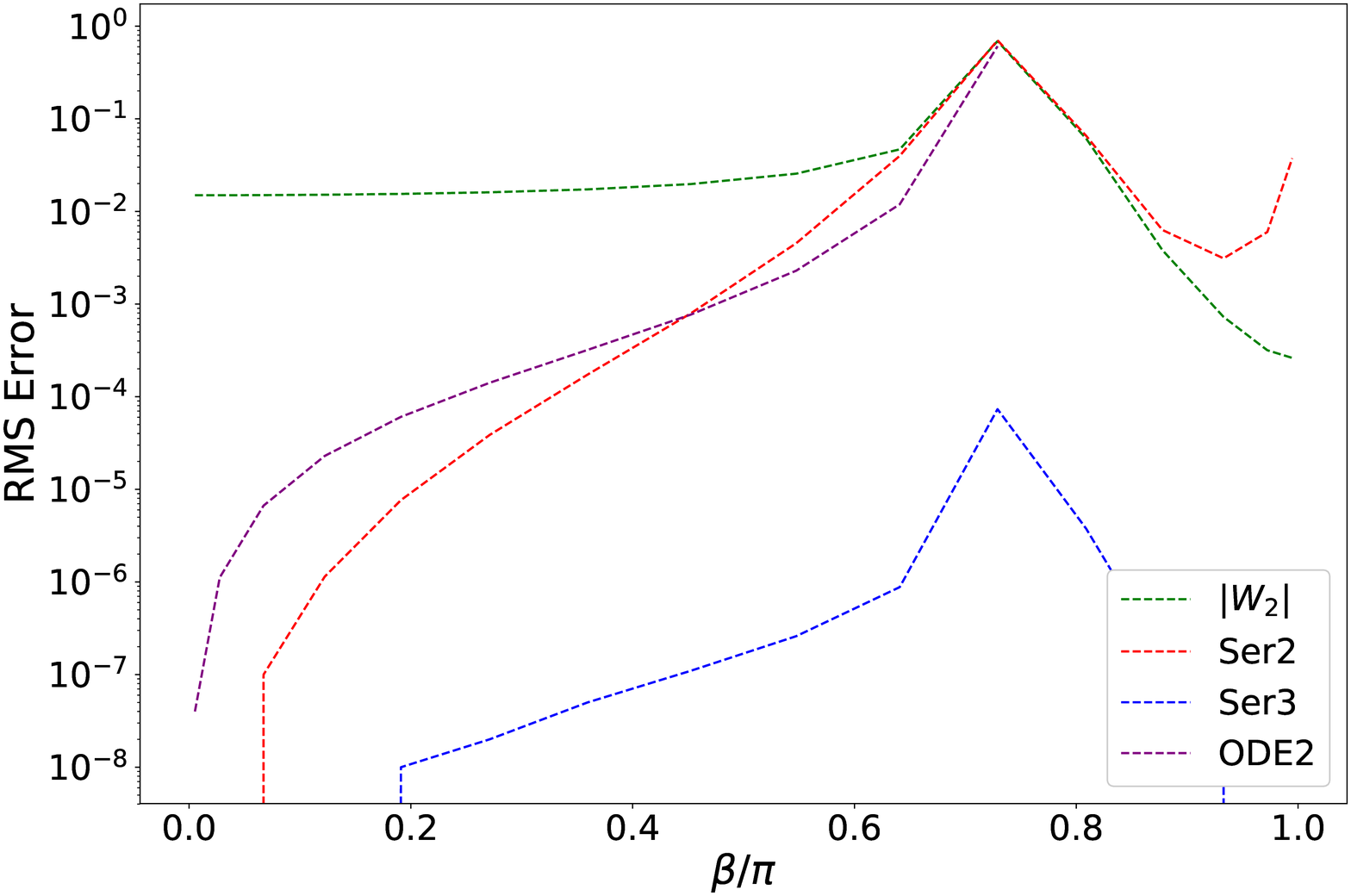}
\caption{RMS error in $W$ from different approximation at $U/t=20$ with exact $U$.   Top panel: Error of $W_0$.  Middle panel: Error of $W_1$. Bottom panel: Error of $W_2$. 
\label{Fig:20PAV}}
\end{figure}

\subsection{Comparison with the ODE Method}
We have seen that the simple power-series expansion approach works well. Here, we want to compare with the earlier ODE-based approach.  

We do not want to say too much about this approach, which we have explained in detail in Ref. \onlinecite{Qiu2017}.  The key is to note that the ODE for $W_n$ depends on $W_{n+1}$ which must be in some way approximated.  The simplest approach for a projected CCSD which relies on $W_0 - W_2$ is to neglect $W_3$ in solving the ODE; this results in the same $\mathcal{O}(N^6)$ scaling as conventional CC, and this method is known as SUCCSD(SD) (ODE2 in this paper).  

To compare everything on even footing, we will this consider the simplest $\mathcal{O}(N^6)$ methods available to us.  In all cases, we retain only $U_1$ and $U_2$, and truncate $W = W_0 + W_1 + W_2$.  The simplest relevant order-by-order expansion uses $n=2$ and $n=3$ to give Ser2 and Ser3, while the ODE analog is ODE2.  Because we have retained only $U_1$ and $U_2$, the result we would like to obtain is PCCSD, but because we have eliminated $W_3$ and higher the best we can expect to do is what we have called Ser.  All calculations in this subsection are conducted in the half-filled 6-site Hubbard model with periodic boundary condition.

As shown in Figure \ref{Fig:HubbardTrunc}, PAV(ODE2) has excellent agreement with PAV(Ser) at small $U/t$ but the two differ substantially as $U/t$ grows (note that PAV(Ser) is indistinguishable from the exact result for small $U/t$).  In contrast, PAV(Ser2) and PAV(Ser3) differ significantly from PAV(Ser) at small $U/t$ but gradually approach it in the more strongly correlated limit.  In this small model problem, in other words, the truncated ODE and truncated series expansion behave very differently when we do not optimize the broken symmetry cluster operator.

On the other hand, the various VAP results are all broadly similar for all $U/t$ with small but non-negligible differences between them.  Only at $U/t \approx 8$ do the various methods differ significantly, where we recall that we have convergence difficulties (and note that here, Ser3 is close to the exact Ser result).

In order to see the difference between different methods in detail, we compare the $W$ amplitudes from every method. Unlike in Ref. \onlinecite{Qiu2017}, we focus on something closer to VAP. First we extract $U_1$ and $U_2$ from VAP(PCCSD) and with these fixed $U$ amplitudes (which we recall are the exact self-consistent $U$ amplitudes in projected CCSD) we calculate $W_0$, $W_1$ and $W_2$ from different approximate methods and compare them to the exact $W$ amplitudes corresponding to the given $U$ amplitudes. Figure \ref{Fig:4PAV} shows that Ser2, Ser3, and ODE2 all show relatively small error at $U/t=4$ but Ser3 performs better at large gauge angle.  However, for large $U/t$ as shown in Fig. \ref{Fig:20PAV}, the situation is somewhat different.  Here, we see that ODE2 and Ser2 have considerable error for large gauge angles, while Ser3 is relatively well behaved.  Taking the $U$ amplitudes from the `Ser' result which optimizes them disregarding $W_3$ and higher yields essentially the same behavior (data not shown).

\begin{figure}
\includegraphics[width=\columnwidth]{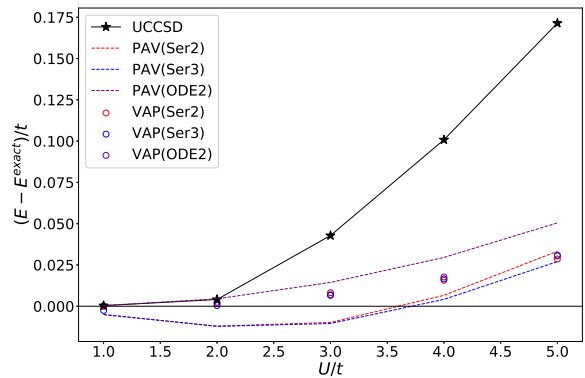} 
\caption{Spin-projected CCSD results in  the 6-electron 10-site Hubbard model.
\label{Fig:DopedHubbard}}
\end{figure}

The ODE- and series expansion approaches seem in a way to be complementary.  When the correlation is not too strong, they perform about equally well.  For stronger correlation atop PHF, where we expect the $U$ amplitudes to be larger, we would expect the series expansion approach to be less successful.  One can see this by considering the doped Hubbard model in Fig. \ref{Fig:DopedHubbard}.  In this case, the PHF method has relative large error and UCCSD is much better. Although for small $U/t$, the two approaches show similar results, our series expansion-based approach fails to converge for $U/t \gtrsim 5$.  The ODE-based approaches, in contrast, work properly.  On the other hand, the ODE-based approach does not readily extend to discrete symmetries where the projection operator is not written as an integral, but our series expansion technique requires no modification for these cases.

\section{Discussion and Conclusion}
\subsection{Discussion}

This work shows an alternative approach to the disentangled cluster approximation, without changing the basic framework established in Ref. \onlinecite{Qiu2017}.  The technique thus shares properties, desirable and undesirable, with this previous work.  It is easy to check that our approach is invariant to occupied-occupied and virtual-virtual rotation.  When the symmetry is not broken, $V_1$ will be zero and the theory goes back to traditional CCSD in which the equations truncate at $\mathcal{O}(U_2^2)$.

On the other hand, the theory is not fully extensive, a drawback shared with projected Hartree-Fock.  Only connected terms in $W$ contribute to the results at each gauge angle, but the final result is a sum over contributions from different gauge angles.  The upshot is that the size extensive component of PHF is the broken-symmetry mean-field energy and presumably the size-extensive component of PCC is the broken-symmetry coupled cluster energy.  Note that the PHF and PCC \textit{wave functions} in the thermodynamic limit differ from their broken-symmetry counterparts, but the energetic difference per particle between the projected and unprojected methods vanishes in the limit of large particle number.

Finally, as has been discussed elsewhere \cite{Qiu2018,Tsuchimochi2018} there is a problem with what we refer to as gauge modes.  Briefly, certain linear combinations of residuals vanish independent of the $U$ amplitudes, simply by virtue of the projection operator.  Essentially, only those components of the broken symmetry cluster operator needed for correlation have physical significance, while those components needed to restore symmetry are rendered irrelevant by the projector.  The practical consequence of this difficulty is that projected CC can be somewhat sensitive to the choice of initial guess; choosing different initial guesses can lead to fluctuations in the energy, as shown in Table \ref{Table:HF}.
\begin{table}
\caption{Total energies in the dissociation of HF (Hartree) with two different initial guesses for the $U$ amplitudes.\label{Table:HF}}

\begin{ruledtabular}
\begin{tabular}{lcc}
  & \multicolumn{2}{c}{Initial Guess}  \\
$R_\mathrm{HF}$ (Bohr)  & UCCSD   & Zero \\
\hline
3.5     &   -100.06248  & -100.06253    \\
5.0     &   -100.03054  &   -100.03057  \\
\end{tabular}
\end{ruledtabular}
\end{table}
While these fluctuations are usually small, they are difficult to eliminate. This linear dependence arising from the gauge modes can also cause convergence difficulties, particularly in the re-coupling region.  It is a good choice to start from some limit in which the solutions are well-defined and use the solution from a previous calculation as a new initial guess to step towards the region in which we are interested. We do not yet have an entirely satisfactory and comprehensive solution to this issue.

\subsection{Conclusion}
Solving strongly-correlated problems requires taking both strong and weak correlations into consideration.  Coupled cluster theory has proven to be an effective theory for weak dynamic correlation, but fails for strongly-correlated problems.  Frequently, however, these strong correlations can be taking into account at the mean-field level through the use of symmetry breaking and restoration.  It seems natural to try to combine PHF with post-Hartree--Fock methods by symmetry-projecting broken-symmetry correlated wave functions, but this is not always straightforward, and is particularly challenging for coupled cluster theory since the exponential nature of the wave function means it contains the full combinatorial number of excitations out of the broken-symmetry mean-field reference despite having a polynomial number of parameters.  In practice one must therefore approximate the projected CC state.

Our preferred disentangled cluster formalism means that the rotated CC wave functions $\mathrm{e}^{W(\theta)} |0\rangle$ to be summed over have the same structure as does the parent broken-symmetry CC state.  That is, they contain all possible excitations, and the cluster operator describing them is intrinsically of the same form as the full CC cluster operator.  By truncating the disentangled cluster operator to some low order, we can readily evaluate the required matrix elements.  In this work, we have shown that one can obtain fairly reliable disentangled cluster operators simply by equating the power series expansion of the rotated coupled cluster wave function to the configuration interaction expansion of the disentangled CC state; in other words, by solving
\begin{equation}
e^{W(\theta)} |0\rangle = R(\theta) \, \sum_{k=0}^n \frac{1}{k!} \, U^k |0\rangle.
\end{equation}
Although this approach might seem to break down when $R(\theta)$ is in some sense large, we have seen that in practice this is not too great a concern, particularly if one optimizes the broken-symmetry cluster operator $U$ in the presence of the projector.

This power series expansion technique complements the ODE-based approach we have used in previous work, and in fact seems to suggest a sort of hybrid technique.  The series expansion gives the (truncated) disentangled cluster operator \textit{exactly} through a given order in $U$, but does not contain any contributions from higher-order terms.  The ODE contains terms from all orders in $U$ but does not give the exact low-order result.  One might be tempted to thus use the series expansion to obtain $W$ through a given order in $U$ and subsequently correct it via the ODE-based formalism to ameliorate the truncation error.

Regardless of how one chooses to approximate the projected CC wave function, energy, and amplitude equations, however, it seems clear to us that projected CC methods have the potential to provide highly accurate results for a wide variety of weakly- and strongly-correlated problems, and the chief difficulty lies in writing sufficiently efficient code and in choosing sufficiently important symmetries to project.

\section{Data Availability}
The data that support the findings of this study are available from the corresponding author upon reasonable request.

\begin{acknowledgments}
This work was supported by the U.S. National Science Foundation under Grant CHE-1762320. G.E.S. is a Welch Foundation Chair (C-0036).  We thank Yiheng Qiu, Guo Chen and Gaurav Harsha for helpful discussion.
\end{acknowledgments}

\bibliography{reference}

\end{document}